\RequirePackage{fix-cm}
\documentclass[a4paper]{article}			 
\usepackage[a4paper]{geometry}
\usepackage{enumitem}
\usepackage{url}
\usepackage{mathtools}
\usepackage{flushend}
\usepackage{color}
\usepackage{graphicx}
\usepackage{epstopdf}
\usepackage{subfig}
\usepackage{balance}
\usepackage{algorithmicx}
\usepackage{algpseudocode}
\usepackage{algorithm}
\usepackage{epsfig,endnotes}
\usepackage{authblk}
\usepackage{enumitem}
\algnewcommand\algorithmicinput{\textbf{Input:}}
\algnewcommand\INPUT{\item[\algorithmicinput]}
\algnewcommand\algorithmicoutput{\textbf{Output:}}
\algnewcommand\OUTPUT{\item[\algorithmicoutput]}
\algnewcommand\algorithmicbegin{\textbf{begin}}
\algnewcommand\BEGIN{\item[\algorithmicbegin]}
\algnewcommand\algorithmicendbegin{\textbf{end}}
\algnewcommand\ENDBEGIN{\item[\algorithmicendbegin]}

\algdef{SE}[DOWHILE]{Do}{doWhile}{\algorithmicdo}[1]{\algorithmicwhile\ #1}%
\newcommand{\overbar}[1]{\mkern 1.5mu\overline{\mkern-1.5mu#1\mkern-1.5mu}\mkern 1.5mu}
\begin{document}

\title{Parallel Computation of PDFs on Big Spatial Data Using Spark}

\author[1]{Ji Liu\thanks{ji.liu@inria.fr}}
\author[2]{Noel Moreno Lemus\thanks{nmlemus@gmail.com}}
\author[1]{Esther Pacitti\thanks{esther.pacitti@lirmm.fr}}
\author[2]{Fabio Porto\thanks{fporto@lncc.br}}
\author[1]{Patrick Valduriez\thanks{patrick.valduriez@inria.fr}}
\affil[1]{Inria and LIRMM, Univ. of Montpelier, France}
\affil[2]{LNCC Petr\'{o}polis, Brazil}



\date{}

\maketitle

\begin{abstract}

We consider big spatial data, which is typically produced in scientific areas such as
geological or seismic interpretation.
The spatial data can be produced by observation (\textit{e.g.} using
sensors or soil instrument) or
numerical simulation programs and correspond to points that represent
a 3D soil cube area.
However, errors in signal processing and modeling create some uncertainty, and thus a lack of accuracy
in identifying geological or seismic phenomenons. Such uncertainty must be carefully analyzed.
To analyze uncertainty, the main solution is to compute a Probability Density Function (PDF) of each point in the spatial cube area.  However, computing PDFs on big spatial data 
can be very time consuming (from several hours to even months on a parallel computer). 
In this paper, we propose a new solution to efficiently compute such PDFs in parallel using Spark,
with three methods: data grouping, machine learning  prediction and sampling.
We evaluate our solution by extensive experiments on different
computer clusters
using big data ranging from hundreds of GB to several TB. The experimental results show that our solution scales up very well and can reduce the execution time by a factor of 33 (in the order of seconds or minutes) compared with a baseline method. 
\end{abstract}

\section{Introduction}

Big spatial data is now routinely  produced and used in scientific areas such as
geological or seismic interpretation \cite{Campisano2016}.
The spatial data are produced by observation, using sensors
\cite{WangL11}, \cite{Chen2014} or soil instruments
\cite{Jackson1999}, or numerical simulation, using mathematical models \cite{Cressie2015}.
These spatial data allow identifying some phenomenon
over a spatial reference \cite{Fotheringham2000}.
For instance,  the spatial reference may be 
a three dimensional soil cube area and the phenomenon a seismic fault,
represented as quantities of interest (QOIs) of sampled points (or points for short) in the cube space.
The cube area is composed of multiple horizontal slices, each slice having multiple lines and each line having multiple points.
A single simulation produces a spatial data set whose points represent a 3D soil cube area.

However, errors in signal processing and modeling create some uncertainty, and thus a lack of accuracy
when identifying phenomenons.
Such uncertainty must be carefully quantified.
In order to understand uncertainty, several simulation runs with different input parameters are usually conducted, thus generating multiple spatial data sets that can be very big,
\textit{e.g.} hundreds of GB or TB.
Within multiple spatial data sets, each point in the cube area is
associated to a set of
different observation values in the spatial data sets. 
The observation values are those observed by sensors, or generated from simulation, at a specific point of the spatial area. 

Uncertainty quantification of spatial data is of much importance
for geological or seismic scientists \cite{Michele2000}, \cite{Trajcevski2011}. It is
the process of quantifying the uncertainty error of each point in the spatial cube space,
which requires computing a Probability Density Function (PDF) of each point \cite{Kathryn2015}. 
The PDF is composed of the distribution type (\textit{e.g.} normal, exponential) and necessary statistical parameters (\textit{e.g.} the mean and standard deviation values for normal and rate for exponential).

\begin{figure}[t]
\centering
\includegraphics[width=8cm]{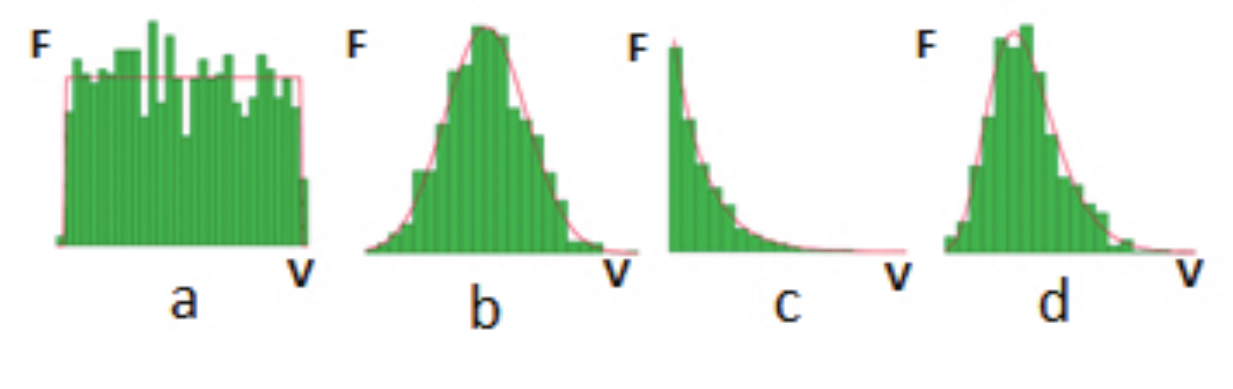}
\caption{\textbf{The distribution of a point.} }
\label{fig:distribution}
\end{figure}

Figure \ref{fig:distribution} shows that the set of observation values at a point may have four distribution types, \textit{i.e.} uniform (a), normal (b),  exponential (c), and log-normal (d). The horizontal axis represents the values (V) and the vertical axis represents the frequency (F). The green bars represent the frequency of the observation values in value intervals and the red outline represents the calculated PDF. 
During the calculation of the PDF at a point, there may be some error between the distribution of the observation values and the calculated PDF.
We denote this error by PDF error (or error for short in the rest of the paper).
In order to precisely fit a PDF based on observation values, we need to reduce this error.
For instance, the set of observation values corresponding to the QOI at a point obeys a normal distribution shown in Figure \ref{fig:distribution} (b). The mean value of the set of observation values (see Equation \ref{eq:mean}) may be used as a representative of QOI since it has the highest chance to be the QOI.
However, the distribution can be different from  normal.
And analyzing uncertainty just using the mean
or standard deviation values is difficult and imprecise.
 For instance, if the distribution type of simulated values is
 exponential (see Figure \ref{fig:distribution} (c)), we should take the
 value zero
(different from the mean value of the simulated values) as the QOI value since it has the highest possibility. 
Once we have the PDF of a point, we can calculate the QOI value that has the highest possibility, with which we can compute the imprecision, \textit{i.e.} quantify uncertainty, of each spatial data set.

\vspace{3mm}

Calculating the PDF that best fits the observation values at each point can be time consuming. For instance, one simulation of an area of 10km (distance) * 10km (depth) * 5km (width) corresponds to 2.4 TB data with 10000 measurements at each point \cite{HPC4E}. This area contains 6.25 * 10$^{8}$ points. The time to calculate the PDF with consideration of 4 distribution types (normal, uniform, exponential and log-normal) can be up to several days or months using a computer cluster.

In this paper, we propose a new solution to efficiently compute PDFs in parallel by taking advantage of Spark
\cite{ZahariaCFSS10}, a popular in-memory big data processing
framework for computer clusters (see \cite{Liu2018}
for a survey on big data systems).
To validate our solution, we use the spatial data generated from simulations
based on the models from the seismic benchmark of the HPC4e project between
Europe and Brazil for oil and gas exploration \cite{HPC4E}.
This benchmark includes models for seismic wave propagation.
In oil and gas exploration, seismic waves
are sent deep into the Earth and allowed to bounce back. Geophysicists
record the waves to learn about oil and gas reservoirs located beneath
Earth's surface. 

The problem we address is how to efficiently compute PDFs under bounded error constraints. In addition to deploying PDF computation over Spark, we propose three methods to efficiently compute PDFs:
data grouping, machine learning (ML) prediction and sampling.
Data grouping consists in grouping similar points to compute the PDF. 
In the original input data, the data corresponding to some points may
be the same or very similar to the data corresponding to a common
point.
ML prediction uses ML classification methods to predict the distribution type of each point. 
Sampling method enables to efficiently compute statistical parameters of a region by sampling a fraction of the total number of points to reduce the computation space.

This paper makes the following contributions:
\begin{itemize}
  \item An architecture to compute PDFs of QOIs in large spatial data sets using Spark.
  \item Three new methods to reduce the time of computing PDFs, \textit{i.e.} data grouping, ML prediction and sampling. 
  \item An extensive experimental evaluation based on the
    implementation of the methods in a Spark/HDFS cluster
and big data sets ranging from 235 GB to 2.4 TB. The experimental results show that our methods scale up very well and reduce the execution time by a factor of 33 (in the order of seconds or minutes) compared with a baseline method.
\end{itemize}

The paper is organized as follows. Section \ref{sec:related} introduces some background on Spark. Section \ref{sec:statement} gives the problem definition.
Section \ref{sec:DP} presents our architecture for computing PDFs with Spark, with its main functions.
Section \ref{sec:solutions} presents our solution 
to compute PDFs in parallel, with three methods, \textit{i.e.} data grouping, ML prediction and sampling. Section \ref{sec:eval} presents our experimental
evaluation on different computer clusters
with different data sizes, ranging from hundreds of GB to several TB.
Section \ref{sec:con} concludes.

\section{Background on Spark}\label{sec:related}

Spark \cite{ZahariaCFSS10} is an Apache open-source data processing
framework. 
It extends the MapReduce model \cite{Dean2004} for two important classes of analytics applications: iterative processing (machine learning, graph processing) and interactive data mining (with R, Excel or Python). 
Compared with MapReduce, it improves the ease of use with the Scala
language (a functional extension of Java) and a rich set of operators (Map, Reduce, Filter, Join, Aggregate, Count, etc.). 
In Spark, there are two types of operations: transformations, which
create a new dataset from an existing one ( \textit{e.g.} Map and
Filter), and actions, which return a value to the user after running a
computation on the dataset (\textit{e.g.} Reduce and Count).
Spark can be deployed on shared-nothing clusters, i.e. clusters of
commodity computers with no sharing of either disk or memory among
computers.
In a Spark cluster, a master node is used to coordinate job execution
while worker nodes are in charge of executing the parallel operations.

Spark provides an important abstraction, called resilient distributed
dataset (RDD),
which is a read-only and fault-tolerant collection of data elements
(represented as key-value pairs)
partitioned across
the nodes of a shared-nothing cluster. 
RDDs can be created from disk-based resident data in files or intermediate data
produced by transformations. They can also be made memory resident
for efficient reuse across parallel operations.

Spark data can be stored in the Hadoop Distributed File System (HDFS)
\cite{Shvachko2010}, a popular open source file system
inspired by Google File System \cite{Ghemawat2003}.
Like GFS, HDFS is a highly scalable, fault-tolerant file system for
shared-nothing clusters.
HDFS partitions files into large blocks, which are distributed and
replicated on multiple nodes. An HDFS file can be 
represented as a Spark RDD and processed in parallel.

Spark provides a functional-style programming interface with various
operations to execute a user-provided function in parallel on an RDD.
We can distinguish between operations without shuffling, \textit{e.g.}
Map and Filter, and with shuffling, \textit{e.g.} Reduce, 
Aggregate and Join.
Shuffling is the process of redistributing the data produced by an
operation, \textit{e.g.} Map,  so that it gets partitioned for the next
operation to be done in parallel, \textit{e.g.} Reduce.
This process is complex and expensive and requires moving data across
cluster nodes.

In this paper, we exploit Spark MLlib \cite{MLib},
a scalable machine learning (ML) library that can handle
big data \cite{Landset2015} in memory by exploiting
Spark RDDs and Spark operations.
In one method which we propose, we use ML techniques to classify the
distribution types of the observation values at different points in order
to reduce useless calculation.

\section{Problem Definition}\label{sec:statement}

A spatial data set contains the information to generate an observation value matrix that corresponds to a three dimensional cube area.
As a result of running multiple simulations, multiple spatial data sets are produced with a set of observation values at each point. 
The set of observation values at each point enables the computation of their mean value, standard deviation values and PDF.

Let us illustrate the problem with the spatial data generated from simulations
based on the models from the seismic benchmark of the HPC4e project \cite{HPC4E}.
An important parameter of the models is 
wave phase velocity, noted $Vp$ in electromagnetic theory, which is the rate at which the phase
of the wave propagates in space.
The models contain 16 layers and each layer is associated to a value of $Vp$.
The top layer delineates the topography and contains the description information of the other 15 layers.
Each of the 15 layers is used to generate the observation values of points in a horizontal space of the cube area.

\begin{figure*}[t]
\centering
\includegraphics[width=14cm]{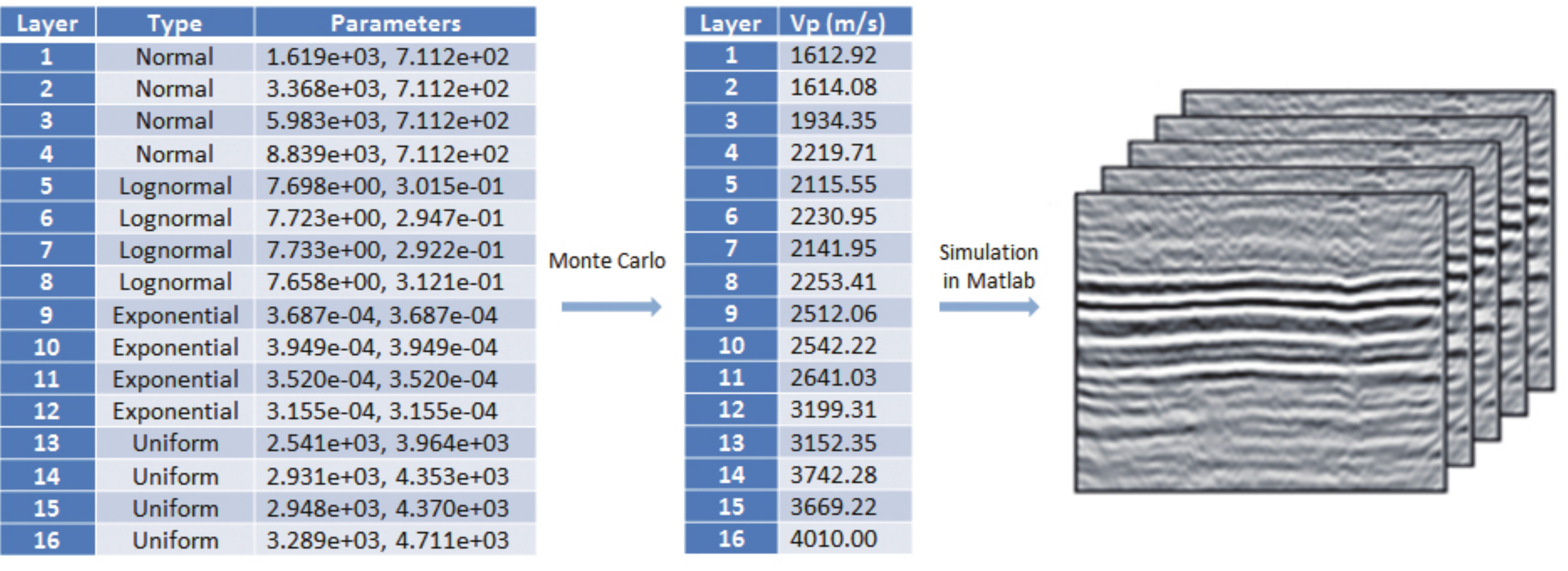}
\caption{\textbf{Data generation from simulation.} This process is repeated multiple times in order to generate multiple data sets. The values of $Vp$ are different at different iterations. $Type$ represents the distribution type. }
\label{fig:datageneration}
\end{figure*}

Since our purpose is to study the uncertainty in the output as a
result of the
wave propagation of the input uncertainty through the model,
we assume that the input value of each layer is uncertain and obeys a PDF.
The distribution type for every four layers are: Normal, Log-normal,
Exponential and Uniform.
We use a Monte Carlo method to generate different sets of the 16 input parameters.
For each set of input parameters, we generate a spatial data set using
the models as shown in Figure \ref{fig:datageneration}.

The main problem we address is to efficiently compute PDFs on big
spatial data sets, such as the seismic simulation data discussed above.
Since it takes much time to compute the PDFs of the points in the
whole cube area, our approach is to divide it
into multiple spatial regions and compute the PDFs of all the points
in a chosen region within a reasonable time.
The spatial region is chosen based on some statistical parameters of the region.
Thus, the main problem is how to efficiently calculate the PDF of each point in a spatial region, \textit{e.g.} a horizontal slice in the cube area, with a small average error between the PDF and the distribution of the observation values.
In order to choose a region, the mean and standard deviation values and the distribution type of a part of the points in the spatial region should be computed. In addition, the average mean value, the average standard deviation value and the percentage of points corresponding to each distribution type in all the points of the region can also be computed. 
We denote these statistical parameters of a spatial region to compute by the features of a spatial region.
A related subproblem is how to efficiently calculate the features of a spatial region.
In this paper, we take a horizontal slice as a spatial region.

Let $DS$ be a set of spatial data sets, $d_k$ be a spatial data set in $DS$ and $N$ be the number of points in a region.
Each point $p_{x,y}$, where $x$ and $y$ are spatial dimensions in the
slice, has a set of values $V = \{v_1, v_2, ..., v_n\}$ while $v_k$ is
the value corresponding to the point $p_{x,y}$ in $d_k \in DS$.
Based on these notations, we define
Equations \ref{eq:mean} - \ref{eq:avgstd} and \ref{eq:averageError}, which
are based on the formulas in \cite{Dixon1968}.
The mean ($\mu_{x,y}$) and standard deviation ($\sigma_{x,y}$) values of a point can be calculated according to Equations \ref{eq:mean} and \ref{eq:std}, respectively.
And the average mean ($\overbar{\mu_{i}}$) and standard deviation
($\overbar{\sigma_{i}}$) values of Slice $i$ can be calculated
according to Equations \ref{eq:avgmean} and \ref{eq:avgstd},
respectively.
The error $e_{x,y,i}$ between the PDF $F$ and the set of observation
values  $V$ can be calculated according to Equation \ref{eq:Error},
which compares the probability of the values in different intervals in
$V$ and the probability computed according to the PDF. The intervals
are obtained by evenly splitting the space between the maximum value
and the minimum value in $V$.
$min$ is the minimum value in $V$, $max$ is the maximum value in $V$ and
$L$ represents the number of all considered intervals, which can be configured. 
$Freq_k$ represents the number of values in $V$ that are in the $k$th interval. 
The integral of $PDF(x)$ computes the probability according to the PDF
in the $k$th interval.
Equation \ref{eq:Error} is inspired by the Kolmogorov-Smirnov Test
\cite{Lopes2011}, which tests whether a PDF is adequate for a data
set. In addition, we assume that the probability of the values outside
the space between the maximum value and the minimum value is
negligible for this equation.
Then, the average error $E$ of Slice $i$ can be calculated according
to Equation \ref{eq:averageError}.

\begin{equation}\label{eq:mean}
\mu_{x,y} = \frac{\sum_{i = 1}^{n} v_i}{n}
\end{equation}

\begin{equation}\label{eq:std}
\sigma_{x,y} = \sqrt{\frac{\sum_{i = 1}^{n} (v_i - \mu)^2}{n - 1}}
\end{equation}

\begin{equation}\label{eq:avgmean}
\overbar{\mu_{i}} = \frac{\sum_{p_{x,y} \in slice_i} \mu_{x,y}}{N}
\end{equation}

\begin{equation}\label{eq:avgstd}
\overbar{\sigma_{i}} = \frac{\sum_{p_{x,y} \in slice_i} \sigma_{x,y}}{N}
\end{equation}

\begin{equation}\label{eq:Error}
e_{x,y,i} = \sum^N_{k = 1}\mid \frac{Freq_k}{N} - \int_{min + (max - min) * \frac{k - 1}{L} }^{min + (max - min) * \frac{k}{L}}PDF(x)dx \mid
\end{equation}

\begin{equation}\label{eq:averageError}
E = \frac{\sum_{p_{x,y} \in slice_i}e_{x,y,i}}{N}
\end{equation}

We can now express the main problem as follows:
given a set of spatial data sets $DS = \{d_1, d_2, ..., d_n\}$ corresponding to the same spatial cube area $C = \{slice_1, slice_2, ..., slice_j\}$, how to efficiently calculate the mean, standard deviation values and the PDF $F$ at each point in $slice_i \in C$ with a small average error $E$ not higher than a predefined average error $\varepsilon$.
In addition, we also need to compute the statistical parameters of slices in order to choose Slice $i$ mentioned in the first subproblem.
Thus, the related subproblem can be expressed as: how to efficiently
calculate the features of a slice when given the same data sets
$DS$. The features are: 
\begin{itemize}
\item $\mu$, $\sigma$ and distribution type of some points in the slice
\item $\overbar{\mu_{i}}$, $\overbar{\sigma_{i}}$ and the percentage of points for each distribution type in all the points of the slice
\end{itemize}

\section{Architecture for Computing PDFs} \label{sec:DP}

\begin{figure}[t]
\centering
\includegraphics[width=8.5cm]{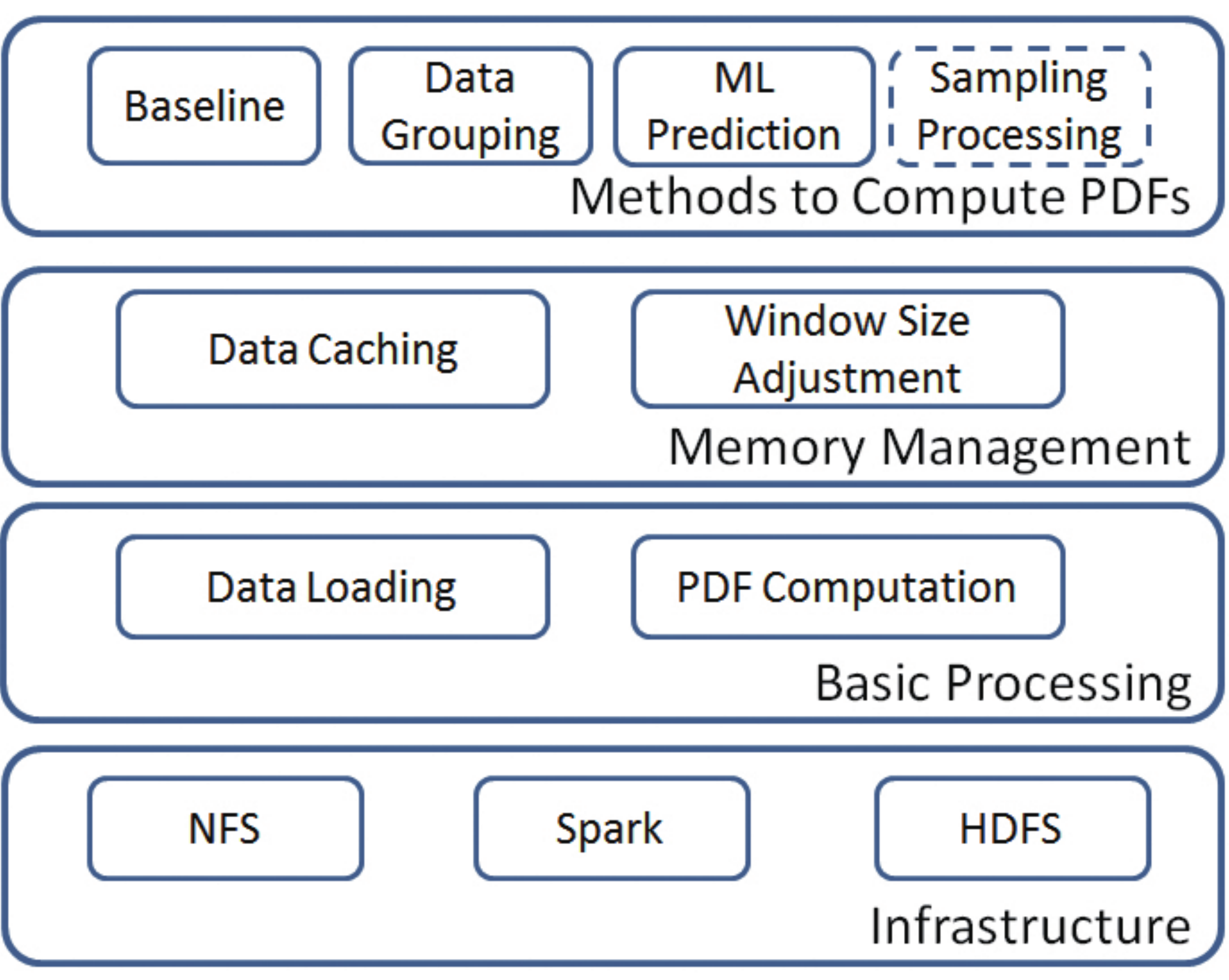}
\caption{\textbf{Architecture for Computing PDFs.}}
\label{fig:archit}
\end{figure}

In this section, we describe the architecture for PDF computation
(see Figure \ref{fig:archit}).
This architecture has four layers,
\textit{i.e.} infrastructure, basic process to compute PDFs,
memory management and methods to compute PDFs. 
The higher layers take advantage of the lower layers' services to
implement their functionality.
The infrastructure layer provides the basic execution environment,
including Spark, HDFS and Network File System (NFS)
in a computer cluster. 
The basic processing layer provides guiding principles to
load the big spatial data and to compute PDFs. 
The memory management layer allows
optimizing the execution of the basic process by caching data
and managing sliding windows on big data.
The methods to compute PDFs are presented in Section \ref{sec:solutions}.

\subsection{\bf{Infrastructure with Spark}} \label{subsec:centralized}

\begin{figure}[t]
\centering
\includegraphics[width=8.5cm]{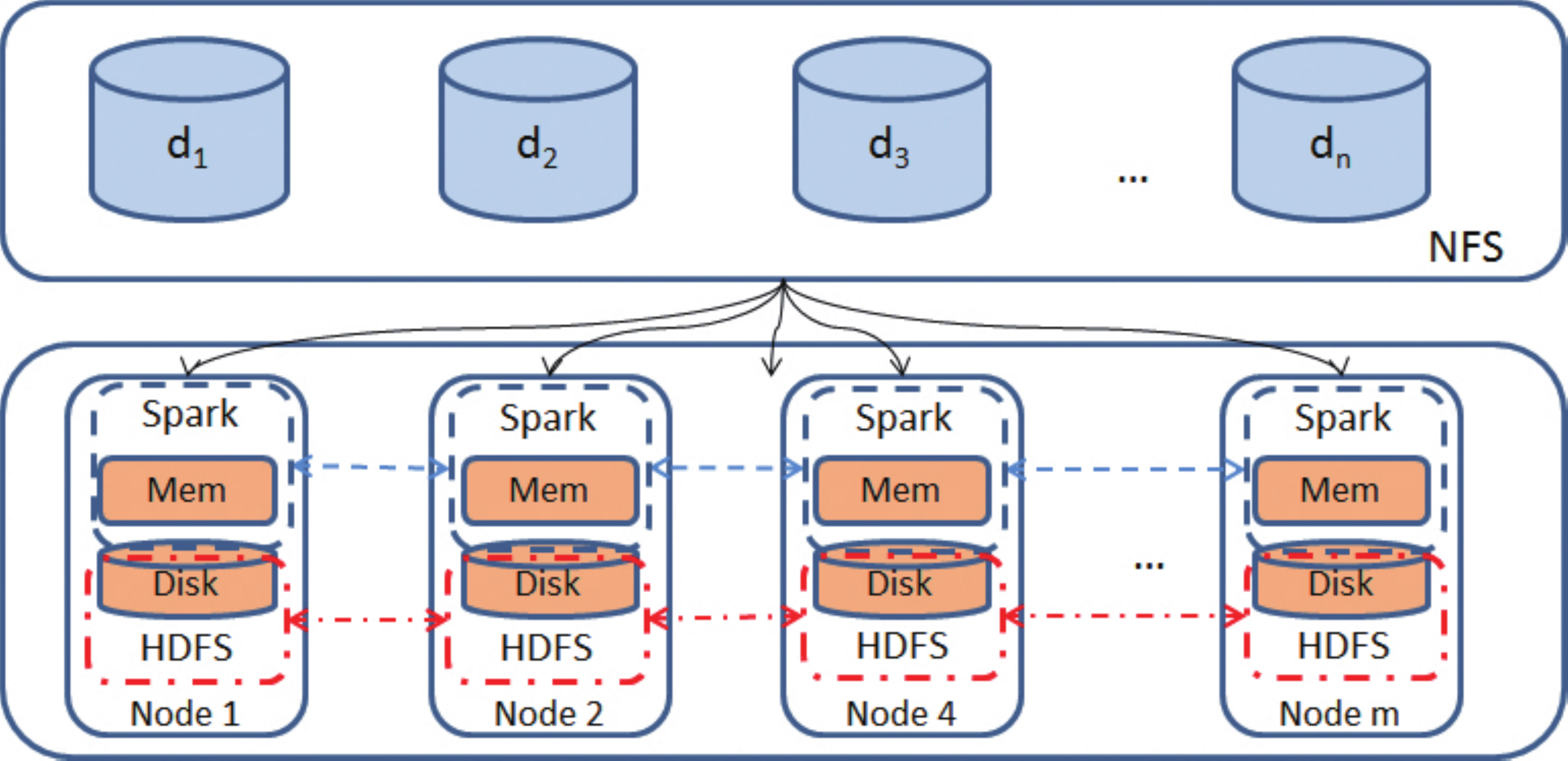}
\caption{\textbf{Infrastructure.}
$d_i$ represents the $i$th spatial data set.}
\label{fig:Inarchit}
\end{figure}

Figure \ref{fig:Inarchit} illustrates the infrastructure we deploy to
process big spatial data.
The big spatial data is produced by simulation application programs
and stored in NFS \cite{nfs}, a shared-disk file system
that is popular in scientific applications.
Spark and HDFS are deployed over the nodes of the computer cluster.
The intermediate data produced during PDF computation and the output
data are stored in HDFS, which provides
persistence and fault-tolerance.

Keeping the input spatial data in NFS allows us to maximize the
use of the cluster resources (disk, memory and CPU), which can
be fully dedicated for PDF computation. With the input data stored in
NFS, the NFS server is
outside the Spark/HDFS cluster and takes care of file management
services, including transferring the data that is read to the cluster
nodes. An alternative solution would
have been to store the input data in HDFS, which would lead to have
HDFS tasks competing with Spark tasks for resource usage on the same
cluster nodes. We did try this solution and it is much less efficient
in terms of data transfer between cluste nodes. This is
because HDFS is more complex and does more work due to
fault-tolerance and data replication.

We use Spark as our execution environment for computing PDFs.
We developed a program written in Scala,
which realizes the functionality of different methods to compute PDFs. 
Once the method to compute PDFs is chosen, the Scala program is
executed as a job within Spark.

\subsection{\bf{Principles for the Basic Processing of PDFs}}\label{sec:workflow}
The basic processing of PDFs consists of data loading, from NFS to
Spark RDDs, followed by PDF computation using Spark.
The data loading process treats the data corresponding to a slice
and pre-processes it, \textit{i.e.} calculates statistical parameters of observation values of each point and relates the observation values of each point to an identification of the point.
The identification of each point is an integer value which represents the location of the point in the cube area.
Then, the PDF computation process groups the data and calculates the
PDFs and errors of all the points in a slice based on the
pre-processed data. 
To make the basic processing of PDFs efficient, we use the
following guiding principles.

\begin{enumerate}
\item \textbf{Parallel data loading. }
To perform data loading in parallel, we store 
the identifications of points in an RDD, which is evenly distributed
on multiple cluster nodes.
For each point in a node, all the corresponding values in different
spatial data sets are retrieved from NFS.
At the same time, the mean and standard deviation values are calculated.
Then, the identification of the point is stored as a key of the point.
The mean and standard deviation values and the observations values are stored as the value of the point.
The key and value of the point are stored as a key-value pair in the RDD.
This loading process is realized by a Map operation in Spark, which is
fully parallel.

\item \textbf{Data grouping. }
In order to avoid repeating the PDF computation for the same or similar sets of observation values, 
we group the data of different points that shares similar statistical features,
\textit{e.g.} the mean and standard values.
Then, a representative point of the group is chosen.
The PDF of the representative point is taken to represent all the points in the group.
Thus, the PDF computation of all the points in the group is reduced to the computation of the representative point.
The grouping can be realized using an \textit{Aggregation} operation in Spark. 
However, the shuffling process in this operation may take much time.
When it takes too much time to group data, we can ignore this principle. 
After data grouping, the set of the identifications of the points in the group is stored as a part of the value in the key-value pair.
For each representative point, the key represents the identification while the value contains the mean and standard deviation values and the observation values.

\item \textbf{Parallel processing of PDFs. }
The PDF of each point (or representative point) is computed based on its observation values.
This process is also realized in a Map operation, which distributes
the key-value pairs in RDD of different points to different nodes.
Once the PDF of a point is calculated, the error between the observation values and the calculated PDF is computed in the same node. If multiple PDFs of different distribution types are computed, the PDF with the minimum error will be chosen as the PDF of the point.
After the parallel processing of PDFs, the key remains the
identification of each point (representative point) while the PDF is
stored as a part of the value in the key-value pair of the RDD. 
In addition, since the mean and standard deviation values and the
observation values
are no longer useful, the corresponding data is removed from the value in the key-value pair.
Finally, the PDF of each point is persisted in a file or database system for future use. In addition, an average error of the PDF of the points in the slice is calculated and shown as the result of executing the Scala program.
 
\item \textbf{Sliding window.} Since aslice can have many points that won't fit
  in memory, we use a sliding window over the slice during
data loading, data grouping and parallel
  processing of PDFs.
A window represents a set of points to process, which correspond to several continuous lines in the slice to process. Any two windows have no intersection. 
After the processing of one window, the process of the next window begins until the end of the slice.
Once the size of the window is configured, it stays the same during execution.
The size of the window has strong impact on execution and must be
chosen carefully (see details in Section \ref{subsubsec:wsa}).

\item \textbf{Use of external programs. }
In the parallel data loading and parallel processing of PDFs, we use
external programs,
which do the specific loading of the points.
Since some Java functions, \textit{e.g.} $skipBytes$ (Skips and discards a
specified number of bytes in the current file input stream),
may not work correctly during the parallel execution of a Map operation in Spark
\cite{Harold2006},
we call an external Java program in the Map operation to retrieve
observation values of a point from different spatial data sets and
pre-process values.
Since the PDF computation is implemented by an external program (in
R), we call it
within the Map operation for the parallel processing of PDFs.
Finally, the output data of the external program is transformed to key-value pairs and stored in RDDs by the same Map operation, which executes the external program.

\end{enumerate}

\subsection{\bf{Memory Management}}

In order to efficiently compute PDFs, we use two memory management 
techniques to optimize the calculation of PDFs over big data: 
data caching and window size adjustment.

\subsubsection{\bf{Data Caching}}

We use data caching, \textit{i.e.} keeping data in main memory, to reduce disk accesses.To identify which data to
cache, we distinguish between four kinds of data: input data,
instruction data, intermediate data and output data. 
The input data is the original data to be processed, \textit{i.e.} the big spatial data
sets. The instruction data is the data corresponding to the external
programs, \textit{e.g.} Java or R programs used in data loading and
PDF computation.
The intermediate data is the data generated by the data loading process or the execution of external programs, and used by subsequent execution.
The output data is the final data generated by the PDF
computation.

We use a simple caching strategy.
We do not cache input data because it can be very big and read only
once. 
We only cache instruction data and intermediate data, which are
accessed much during execution. However, intermediate data that is not
used in subsequent operations is removed from main memory.
The output data is written to memory first and then persisted.

During execution, some intermediate data that is stored in Spark RDDs
can be cached using the Spark $Cache$ operation, which stores RDD data
in main memory.
However, the instruction data of external programs and the
intermediate data directly generated by executing these external
programs are outside of RDDs.
We cache this data in temporary files in memory, using a
memory-based file system \cite{Snyder90}.
Then, the information in the cached files is retrieved and stored as
intermediate data in RDDs, which can be again cached using the $Cache$
operation of Spark.

\subsubsection{\bf{Window Size Adjustment}}
\label{subsubsec:wsa}

The size of the sliding window is critical for efficient PDF
computation.
When it is too small, the degree of parallel execution among multiple
cluster nodes is low.
Increasing the window size increases the degree of parallelism, but
may introduce some overhead in terms of data transfers among nodes or
management of concurrent tasks.
Thus, it is important to find an optimal window size in order to make a good trade-off between the
degree of parallelism and overhead.

To find the optimal window size, we distinguish between the data
loading process and the PDF computation process, which have different data
access patterns.
In data loading, the processing of each point can be done independently
by a Map operation in parallel.
Thus, the overhead of data transfers and concurrent task management with a big window size is small.
Thus, we can choose a window size that ensures that there is enough work to do for
each node and that multiple nodes can be used.
For instance, consider a computer cluster with $n$ nodes, each node with $c$ CPU cores. 
Assuming that the data loading for each point can occupy a CPU core,
we can choose a maximum window size corresponding to $n$*$c$ 
points. If the number of points is less than $n$*$c$, we choose the maximum number of points as the window size.

During PDF computation, the overhead of having a big window can be high since the processing of different points are not independent especially when we use data grouping. 
For instance, when the size of window is bigger, the data of more points is present at node. 
In order to group data, the data of each point need to be compared with the data of more points, which takes more time. 
In addition, there is much more data transferred among different nodes in the data shuffling process of data grouping.
Furthermore, since the PDF computation takes much time, the
management of concurrent tasks within each node also increases
execution time for a big window.
As a result, the overhead of having a big window becomes high for PDF computation.

To find an optimal window size, we test the Scala
program on a small workload (with a small number of points)
with different window sizes, and then use
the optimal size for the PDF computation of all the points in the slice.

\section{Methods to Compute PDFs}\label{sec:solutions}
In this section, we present our methods to compute PDFs efficiently.
First, we introduce a baseline method, which will be useful for
comparison.
Then, we propose two methods, \textit{i.e.} data grouping and ML
prediction, to compute PDFs, which addresses the
main problem defined in Section \ref{sec:statement}. 
Finally, we propose a sampling method to calculate the features of a
slice, which addresses the related problem.

\subsection{\bf{Baseline Method}}\label{sec:multisite}

The baseline method
computes the PDF of each point in a slice as follows (see Algorithm \ref{alg:AP}).
Line 2 loads the spatial data sets and calculates the mean and
standard deviation values of each point by using Algorithm
\ref{alg:DL}. Lines 3 - 13 compute the PDF for all the points in the
$i$th slice. Line 5 gets a window in the slice. Line 6 selects all
the points in the window to process.
For each point in the window, Line 8 computes the PDF based on the
observation values and the error between the PDF and the
observation values.
This can be achieved by executing an R program. 
The loop of Lines 7-10 can be executed in parallel using the Map operation in Spark.
The $ComputePDF\&Error$ function is realized by Algorithm
\ref{alg:Ana}.
The data is persisted in the storage resources (Line 11) and the
average error $E$ is calculated (Line 14). Lines 3-14 correspond
to the PDF computation process.

\begin{algorithm}
\caption{PDF computation}\label{alg:AP}
\begin{algorithmic}[1]
\INPUT $DS$: a set of spatial data sets corresponding to a spatial cube area; 
$i$: the $i$th slice to analyze
\OUTPUT $PDF$: the PDF of all the points in the $i$th slice of the
cube area; $E$: the average error between the PDF and the observation
values of all the points in the $i$th slice
\State $PDF\gets \emptyset$
\State $RawData\gets loadData(DS, i)$
\While{not all points in $slice_i$ are processed}
\State $PDF\gets \emptyset$
\State $window\gets GetNextWindow( slice_i )$  
\State $Points\gets Select(window, RawData)$
\ForAll{$p \in Points$}
\State $(pdf, error)\gets ComputePDF\&Error(p, RawData)$
\State $pdfs\gets pdf \cup wsf$
\EndFor
\State persist($pdf$)
\State $PDF\gets PDF \cup pdfs$
\EndWhile
\State $E\gets Average(error)$
\ENDBEGIN
\end{algorithmic}
\end{algorithm}

Algorithm \ref{alg:DL} loads and preprocesses the spatial data. Line 4
chooses a window to load the data. Then, for each point in the window,
the data in each data set of $DS$ is loaded (Lines 6-10). This
process is realized in a Map function in Spark. A Java
program is called in the Map function to read
the data at a specific position instead of loading all the data. Then,
the mean and standard deviation
values are calculated (Lines 11 and 12). Finally, the loaded data is
cached in memory in a Spark RDD (Line 16).

\begin{algorithm}
\caption{Data loading}\label{alg:DL}
\begin{algorithmic}[1]
\INPUT $DS$: a set of data sets corresponding to a spatial cube area; 
$i$: the $i$th slice to analyze
\OUTPUT $RawData$: mean, standard deviation and the original data set
of each point in the cube area
\State $RawData\gets \emptyset$
\While{$RawData$ does not contain all the points in $slice_i$}
\State $windowData\gets \emptyset$
\State $window\gets GetNextWindow( slice_i )$  
\ForAll{$p \in window$}
\State $rd\gets \emptyset$
\ForAll{$ds \in DS$}
\State $data\gets GetData(ds, p, i)$
\State $rd\gets data \cup rd$
\EndFor
\State $\mu\gets ComputeMean(rd)$
\State $\sigma\gets ComputeStd(rd)$
\State $rd\gets \mu \cup \sigma \cup rd$
\State $windowData\gets rd \cup windowData$
\EndFor
\State $Cache(windowData)$
\State $RawData\gets windowData \cup RawData$
\EndWhile
\ENDBEGIN
\end{algorithmic}
\end{algorithm}

Algorithm \ref{alg:Ana} computes the PDF of a point with the smallest
error in a set of candidate distribution types. Line 3 calculates
the statistical parameters of PDFs based on different distribution
types in a set of distribution candidates $Types$. For instance, the
parameters for normal are mean and standard deviation values while the
parameter for exponential is rate. The more types are considered in
$Types$, the longer the execution time of Algorithm \ref{alg:Ana}
is. Then, the corresponding error of the PDF based on $type$ and
$parameters$ are calculated using Equation \ref{eq:Error}. Finally, the
PDF that incurs the smallest error is chosen (Line 7).

\begin{algorithm}
\caption{PDF computing}\label{alg:Ana}
\begin{algorithmic}[1]
\INPUT $d$: a set of observation values for a point; $Types$: a set of
distribution types
\OUTPUT $PDF$: the PDF of the point; $error$: the error between the
PDF defined by the distribution type and the statistical parameters
and the observation values of the point
\State $results\gets \emptyset$
\ForAll{$type \in Types$}
\State $parameters\gets fitDistribution(d, type)$
\State $pError\gets CalculateError(type, parameters, d)$ \Comment{According to Equation \ref{eq:Error}}
\State $results\gets \{(type, parameter, pError)\} \cup results$
\EndFor
\State $(PDF, error)\gets GetSmallestError(results)$
\ENDBEGIN
\end{algorithmic}
\end{algorithm}

We exploit Algorithms \ref{alg:AP} and \ref{alg:DL} in both the data
grouping and ML prediction methods. However, the
$Select$ and $ComputePDF\&Error$ functions are different in different methods.

\subsection{\bf{Data Grouping}}
\label{subsec:DG}

Some points may have the same distribution of
observation values. Thus, using the data grouping method, we can execute the PDF computation process only
once and use the result to represent all the corresponding
points. In this method, the
$Select$ function in Algorithm \ref{alg:AP} has two steps. First, the
points with exactly the same mean and standard deviation values are
aggregated into a group. The grouping can be realized by the
\textit{aggregation} operation in Spark. Then, one point in each group
is selected to represent the group of points. Then, the data
corresponding to selected points are processed to compute the PDFs.

In some cases, although some points may share the same PDF, the
observation values of the points are slightly different. And the mean
and standard values may different by a very small fluctuation. In
this case, we can cluster the points
that have similar mean and standard deviation values with an
acceptable error.

When the number of nodes in the cluster is small and the window size is
small, there is few data to be transferred for the grouping. And
the grouping method can avoid repeated execution on the same set of
observation values. However, when the number of nodes is high or the
window size is big, there is much more data to be transferred among
different nodes, which may take much time. In some situations, the
time to transfer the data may be longer than the time to run the
repeated execution. In addition, if a point corresponds to big amounts
of data (many observation values), even though the window size is small,
the shuffling process of data grouping may also take much time. In
this case, the data grouping method is not efficient.

\subsubsection{\bf{Reuse Optimization}}

When there are points of the same mean and standard
deviation values in different windows, we can reuse the existing
calculated results in order to avoid new executions. 
Thus, we propose
the reuse optimization method, which not only aggregates the data to
groups but also checks if there are already existing results,
\textit{i.e.} the PDFs for the point of the same mean and standard
deviation values, generated from the previous execution of processed
windows.
We expect this method to be better than data grouping only. However,
it may take time to store all the calculated results and to search
existing PDFs from a large list of previously generated results. This
extra time may be longer than the reduced time, \textit{i.e.} the time
to compute the PDF on the data set (the execution time of Algorithm
\ref{alg:Ana}).

\subsection{\bf{ML Prediction}}
\label{subsec:MLP}

In this section, we propose an ML prediction method based on a decision
tree to compute PDFs.
Decision tree classifier \cite{Safavian1991} is a typical ML technique
to classify an object into different categories. The basic idea is to
break up a complex decision into a union of several simpler decisions,
hoping the final solution obtained this way would resemble the
intended desired solution. The decision tree can be described as a
tree selected from a lattice \cite{BelohlavekBOV09}. A tree $G = (V,
E)$ consists of a finite, nonempty set of nodes $V$ and a set of edges
$E$. A path in a tree is a sequence of edges. If there is a path from
$v$ to $w$ ($v≠w$), then $v$ is a proper ancestor of $w$, $w$ is a
proper descendant of $v$ and the path contains an entry edge for $w$
and an out edge for $v$.

\begin{figure}[t]
\centering
\includegraphics[height=6.5cm]{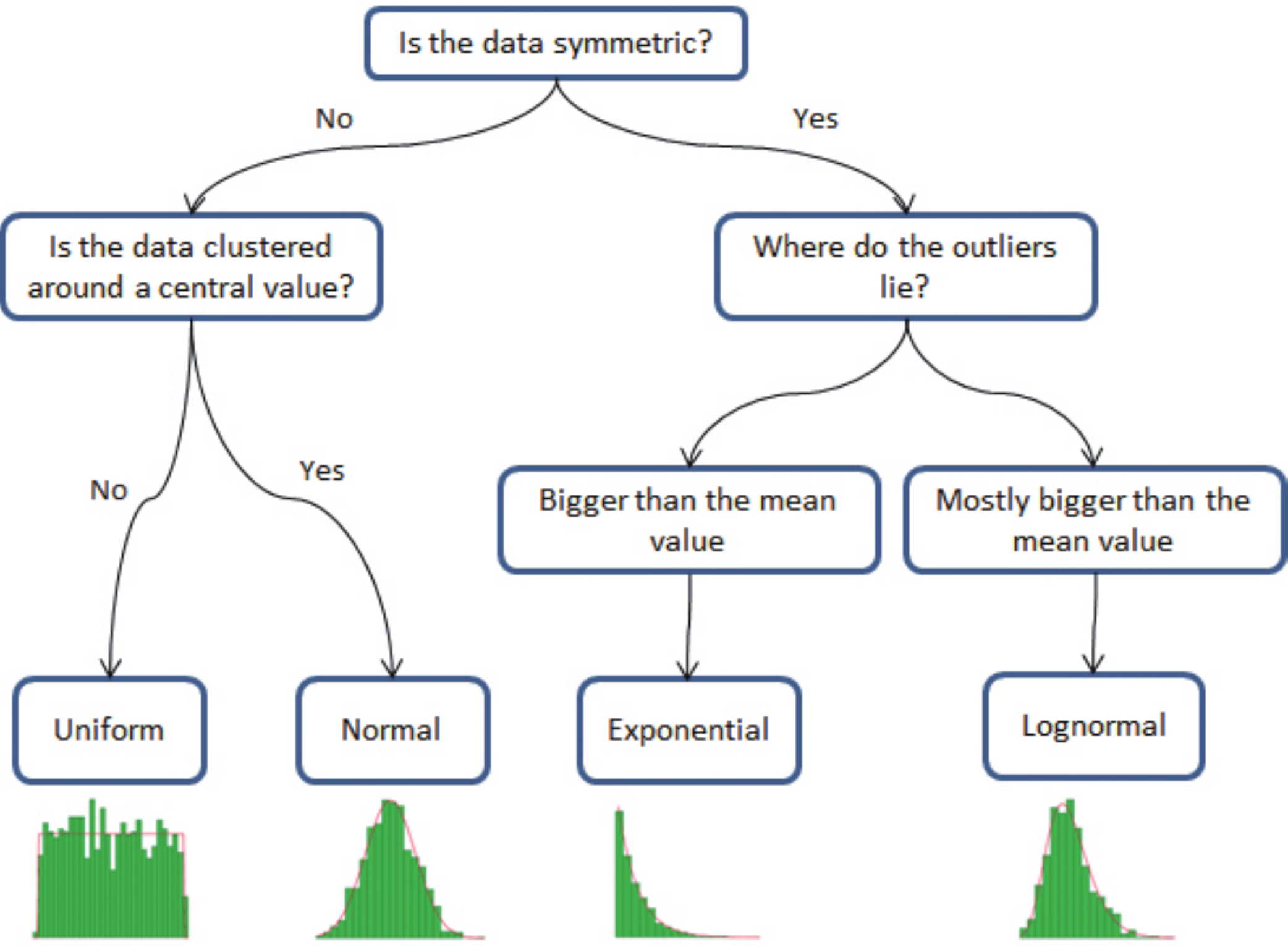}
\caption{A decision tree to choose the distribution type for a set of data.}
\label{fig:statis}
\end{figure}

A tree has only one root node that does not have any entry edge and all
the other nodes have only one entry edge.  Each node has only one path
from the root. The depth of the graph in a tree is the length of the
largest path from the root to a leaf. A leaf is the node that has no
proper descendant. The data to be used to generate a decision tree is
training data. In order to generate a decision tree, the training data
can be split into different data sets (bins) according to different
features and each data set is a unit to be processed. The maximum
number of bins is defined in order to reduce the time to generate a
decision tree. Figure \ref{fig:statis} shows a decision tree of
depth 3 to choose a distribution type for a data set. However,
in this paper, we take some statistical features as parameters to
choose a distribution as explained below.

In Algorithm \ref{alg:Ana}, the execution of Lines 3-5 is repeated
several times (the number of distribution type candidates), which is
very inefficient. We assume that we can learn the correlation among
statistical features, \textit{e.g.} mean and standard deviation
values, and the distribution type. Then, we can directly predict the
distribution type based on the relationship and use the predicted
distribution type to execute Lines 3-5 in Algorithm \ref{alg:Ana}
once for each point.

We assume that we have some previously generated output data, which
contains the results of several points, \textit{i.e.} the mean and standard
deviation values and the type of distribution.
Before execution of our method, we can generate a ML model (that
correlates between statistical features and distribution types),
\textit{i.e.} decision tree, based on the previous existing
data. Then, we use Algorithm \ref{alg:ML} to replace Algorithm
\ref{alg:Ana}.

When the points in the current slice have the
same correlation between the statistical features and the distribution
types as that in the previously generated output data, we can use ML
prediction. Generally, 
the points in different slices but corresponding to the same spatial data set
have the same correlation. Thus, we can use the
output data generated based on some points in one slice,
\textit{e.g.} Slice 0, to generate decision tree model and use the
model to calculate PDFs of the points in other slices, \textit{e.g.}
Slice 201.

In addition, since the ML approach optimizes the $ComputePDF\&Error$ 
function, it can be combined with other methods.
Thus, we call the
pure adoption of ML: ML or baseline + ML; the combination 
of data grouping and ML: data grouping + ML or grouping + ML; the 
combination of reuse and ML: reuse + ML.
Grouping + ML first 
uses the data grouping methods to group points and then use ML prediction
to calculate PDF and error for each representative point. Reuse + ML 
first groups the points using the data grouping method and then
searches the PDF corresponding to the set of the mean and standard deviation 
values of each representative point in the results generated by previous 
execution. If there is no results for the set of mean and standard deviation 
values, it uses the ML method to calculate the PDF and the error.

\begin{algorithm}
\caption{PDF computing based on ML prediction}\label{alg:ML}
\begin{algorithmic}[1]
\INPUT $d$: a vector of
observation values for a point; $model$: the decision tree; $\mu$: the mean value of $d$; $\sigma$: the standard deviation value of $d$
\OUTPUT $T$: the distribution type of the point; $P$: the parameters
of the distribution; $error$: the error between the PDF defined by the
distribution type and the parameters of the distribution and the real
distribution of data in $d$
\State $T\gets predict(model, \mu, \sigma)$
\State $P\gets fitDistribution(d, type)$
\State $error\gets CalculateError(type, parameter, d)$ \Comment{According to Equation \ref{eq:Error}}
\ENDBEGIN
\end{algorithmic}
\end{algorithm}

\subsubsection{\bf{ML Model Generation}}
\label{subsubsec:MLMG}

In order to generate the decision tree model, there are some
hyper-meters to determine, \textit{e.g.} the depth of the tree
($depth$) and the maximum number of bins ($maxBins$). In order to tune
the hyper-meters, we randomly split the previously generated output data into two data
sets, \textit{i.e.} training set and validation set. We use the
training set to train the model with different combinations of
hyper-meters. Then, we test the generated models on the validation set
in order to choose an optimal combination of hyper-meters. We take
the wrong prediction rate in the validation set as the model error while
tuning the hyper-meters. The model error represents the preciseness of
the decision tree model.

The model error decreases at the beginning
and then increases, because of overfitting \cite{Zadrozny2001}, as the
values of $depth$ and $maxBins$ increase. While the time to train the
model becomes longer when the values of $depth$ and $maxBins$
increase. We choose the minimum values of $depth$ and $maxBins$, from
which the error does not decrease when they ($depth$ or $maxBin$)
increase. The process to tune the hyper-meters can take much time. We
assume that the points in different slices have the same correlation
between the statistical features and distribution types. We also assume that the
hyper-meters can be shared to train the models based on different previously generated output data in order to
avoid tuning the hyper-meters and to reduce the time to
train the decision tree model. Thus, the chosen values of $depth$ and
$maxBins$ are used as fixed hyper-meters to generate the decision tree
model for different previously generated output data.

Within the previously generated output data, each point has mean and standard deviation
values and the distribution type.
Since we have fixed hyper-meters, we do not need to use a validation
data set to tune the hyper-meters.
Thus, in order to generate the model, we randomly partition the
previously generated output data set into two parts, \textit{i.e.} training set and test
set. In the training processing, we train the model using the training
set. The input of the trained model is the mean and standard values
and the output is a predicted distribution type. After generating the
model, we take the wrong prediction rate in the test set as the model
error while training models. During the PDF computation process, the
generated model is broadcast to all the nodes, which reduces communication cost.
.

There are scenarios in which the mean and standard deviation share the
same values, respectively, but the distribution type is different. In
this situation, we can take into consideration other normalized
moments \footnote{The $n$th moment is defined as:
\begin{equation}\label{eq:moment}
\mu_n = \sum_{i = 1}^m{(x_i - \mu)^n}
\end{equation}
where $m$ is the number of values in the data set, $x_i$ is the $i$th
value in the data set and $\mu$ is the mean value of the data set.
}, \textit{e.g.} 3th, 4th etc. However, it may take additional time to
calculate other normalized moments. Thus, we only consider the mean
and standard deviations. We will experiment with data
sets where the points of the same set of mean and standard deviation
values have the same distribution types.

\subsection{\bf{Sampling}}\label{subsec:fproc}

In order to choose a slice to compute PDFs, we need to compute the
features of a slice very quickly. We propose a sampling method
(see Algorithm \ref{alg:FP}) that samples the points and uses
ML prediction to generate the distribution type of each point. This
method does not run the statistical calculation of the each point in the
$ComputePDF\&Error$ function in Algorithm \ref{alg:AP} in order to
reduce execution time.

In Algorithm \ref{alg:FP}, Line 2 samples the points in slice $i$
based on a predefined sampling rate. A sampling rate represents the
ratio between the selected points from the sampling process and all
the original points. Lines 4-14 implement the process
that loads the data from the data sets and calculates the mean and
standard deviation values for each double sampled point. Line 15
groups the data as explained in Section \ref{subsec:DG}. When the
number of nodes in the cluster is high, we can remove Line 15 in
order to reduce the time of shuffling. Lines 17 -20 calculate the
distribution type of each double sampled point based on a decision tree
(see Section \ref{subsubsec:MLMG} for details). Lines 22-26 calculate
the final results, \textit{i.e.} the average mean value, the average
standard deviation and the percentage of different distribution
types. Lines 17 - 26 correspond to the PDF computation process.

\begin{algorithm}
\caption{Sampling process}\label{alg:FP}
\begin{algorithmic}[1]
\INPUT $DS$: a set of data sets produced by simulations; 
$i$: the $i$th slice to analyze;
$model$: the decision tree corresponding to the relationship between the mean and standard value and the distribution type;
$Types$: a set of distribution types;
$rate$: the sampling rate
\OUTPUT $\overbar{\mu}$: the average mean value of the slice; $\overbar{\sigma}$: the average standard deviation of the slice; $TypesPercentage$: the percentage of distribution types of the points in the slice
\State $SF\gets \emptyset$
\State $Points\gets Sample(slice_i, rate)$
\State $RawData\gets \emptyset$
\ForAll{$p \in points$}
\State $rd\gets \emptyset$
\ForAll{$ds \in DS$}
\State $data\gets GetData(ds, p, i)$
\State $rd\gets data \cup rd$
\EndFor
\State $\mu\gets ComputeMean(rd)$
\State $\sigma\gets ComputeStd(rd)$
\State $rd\gets \mu \cup \sigma \cup rd$
\State $RawData\gets rd \cup RawData$
\EndFor
\State $Points\gets selectByGrouping(RawData)$
\State $allTypes\gets \emptyset$
\ForAll{$p \in Points$}
\State $type\gets predict(model, RawData)$
\State $allTypes\gets type\cup allTypes$
\EndFor
\State $TypesPercentage\gets \emptyset$
\ForAll{$type \in Types$}
\State $pct\gets calculatePercentage(type, allTypes)$
\State $TypesPercentage\gets pct\cup TypesPercentage$
\EndFor
\State $(\overbar{\mu}, \overbar{\sigma})\gets averageCalculation(RawData)$
\ENDBEGIN
\end{algorithmic}
\end{algorithm}

We can randomly sample the points. The random sampling method takes
very time while the selected points may contain little repeated
information. We could also use a k-means clustering algorithm \cite{Meng2016}
and choose the point that is the closest to the center of
each cluster as double sampled points.
The double sampled data
generated by k-means contains diverse information,
which does not help much to choose a slice (see details in Section \ref{subsec:EFP}).
Furthermore, k-means may
take much time to converge. Therefore, instead of k-means,
we use random sampling in Algorithm \ref{alg:FP}.

\section{Experimental Evaluation}\label{sec:eval} 

In this section, we evaluate and compare the different methods
presented in Section \ref{sec:solutions} for different data sets and
cluster sizes. To ease reading, we call each method by a name as:
Baseline, Grouping, Reuse, ML and Sampling.
First, we introduce the experimental setup, with two
different computer clusters (a small one and a big one). Then, we
perform experiments with a 235 GB data set. Finally, we perform
experiments with big data sets (1.9 TB and 2.4 TB).

\subsection{\bf{Experimental Setup}}
\label{subsec:ES}

In this section, we present the different spatial data sets, the
candidate distribution types (see Algorithms
\ref{alg:Ana} and \ref{alg:FP}) and the cluster testbeds, which we use
in our various experiments.

To generate the spatial data, we use the  UQlab framework \cite{UQdoc} to produce
16 values as the input parameters, \textit{i.e.}
$Vp$, of the models from the seismic benchmark of the HPC4e project \cite{HPC4E}.
The input parameters obey four distribution types, \textit{i.e.} normal, exponential, uniform and log-normal. 
In each simulation, we generate a set of the input parameters
according to the PDF of each layer and generate a spatial data set by
using the set of the input parameters and the models.
We run the simulation multiple times and generate three sets of
spatial data sets, denoted by
$Set1$, $Set2$ and $Set3$.
The simulation is repeated 1000 times to generate $Set1$.
$Set1$ contains 1000 files, each of which is a spatial data set and has 235 GB.
In $Set1$, the dimension of the cube area is 251 * 501 * 501, \textit{i.e.} 501 slices, each slice has 501 lines and each line is composed of 251 points. 
To generate $Set2$, we run the simulation 1000 times while the dimension is 501 * 1001 * 1001.
$Set2$ has 1.9 TB.
A point is associated to 1000 observation values in both $Set1$ and $Set2$. 
Finally, we run the simulation 10000 times with the dimension of
251 * 501 * 501 in order to generate $Set3$, which has 2.4
TB. In this data set, a point is associated to 10000 observation
values.
We use the same slice (Slice 201 because it has interesting information)
in all the experiments.
We consider two sets of candidate distribution types $Types$, introduced
in Algorithms \ref{alg:Ana} and \ref{alg:FP}.
The first set is the set of distribution types of the input
parameters, \textit{i.e.} normal, exponential, uniform and log-normal,
since we assume that the distribution types of different points are
within those of the input parameters of the simulation.
In addition, we assume that the distribution types may belong to other
types beyond the scope of the distribution types of the input
parameters because of non-linear relationship between the input
parameters and the values of each point of the spatial cube area. With
this assumption, we have a second type of candidate distribution
types, \textit{i.e.} normal, exponential, uniform, log-normal, Cauchy,
gamma, geometric, logistic,
Student's t and Weibull.
We call the first set $4-types$ and the second $10-types$.

We use two cluster testbeds, each with NFS, HDFS and Spark deployed.
The first one, which we call LNCC cluster, is a
cluster located at LNCC with 6 nodes, each having
32 CPU cores and 94 GB memory. The second one
is a cluster of Grid5000, which we call
G5k cluster, with 64 nodes, each 
having 16 CPU cores and 131 GB of RAM storage.

\subsection{\bf{Experiments on a 235 GB Data Set}}
\label{subsec:EBSD}

In this section, we compare the different methods based on
the spatial data set $Set1$ of 235 GB.
First, we compare the performance of different methods using the LNCC
cluster:
Baseline, Grouping, Reuse and the
combination of these methods with ML.
Then, we study the scalability of different methods using the G5k
cluster.
Finally, we study the performance of Sampling.

We take the relationship between the set of mean and standard
deviation values and the distribution types of 25000 points in Slice
0 in the previously generated output data of $Set1$ to tune the hyper-meters of the decision
tree. The hyper-meters are tuned at the very beginning, which takes
18 minutes for $4-types$ and 22 minutes for $10-types$ using Spark
on a workstation with 8 CPU cores and 32 GB of
RAM.

We take advantage of the
previously generated output data of $Set1$ to generate the decision
tree model for the experiments of this section.  The model error is
0.03 for $4-types$, and 0.08 for $10-types$. Although the model
error of $10-types$ is bigger than that of $4-types$
, the average error $E$ in Algorithm
\ref{alg:AP} is smaller. The time to train the model ranges from 1
to 20 seconds, which is negligible compared with the
execution time of the whole PDF computation process.

\subsubsection{\bf{Performance Comparisons}}
\label{subsec:CFN}

In this section, we compare the performance
of different methods, \textit{i.e.} Baseline, Grouping, Reuse, and the
combination with ML, using the LNCC cluster.
First, we execute the Scala program (for computing PDFs with different
methods) on a small workload (6 lines and 3006 points). Second, we compare
the performance of different methods for different window sizes.
Finally, we compare the performance of
different methods with the tuned window size for the whole slice.

\paragraph{Experiments with Small Workload}

\begin{figure}[t]
\centering
\includegraphics[width=8.5cm]{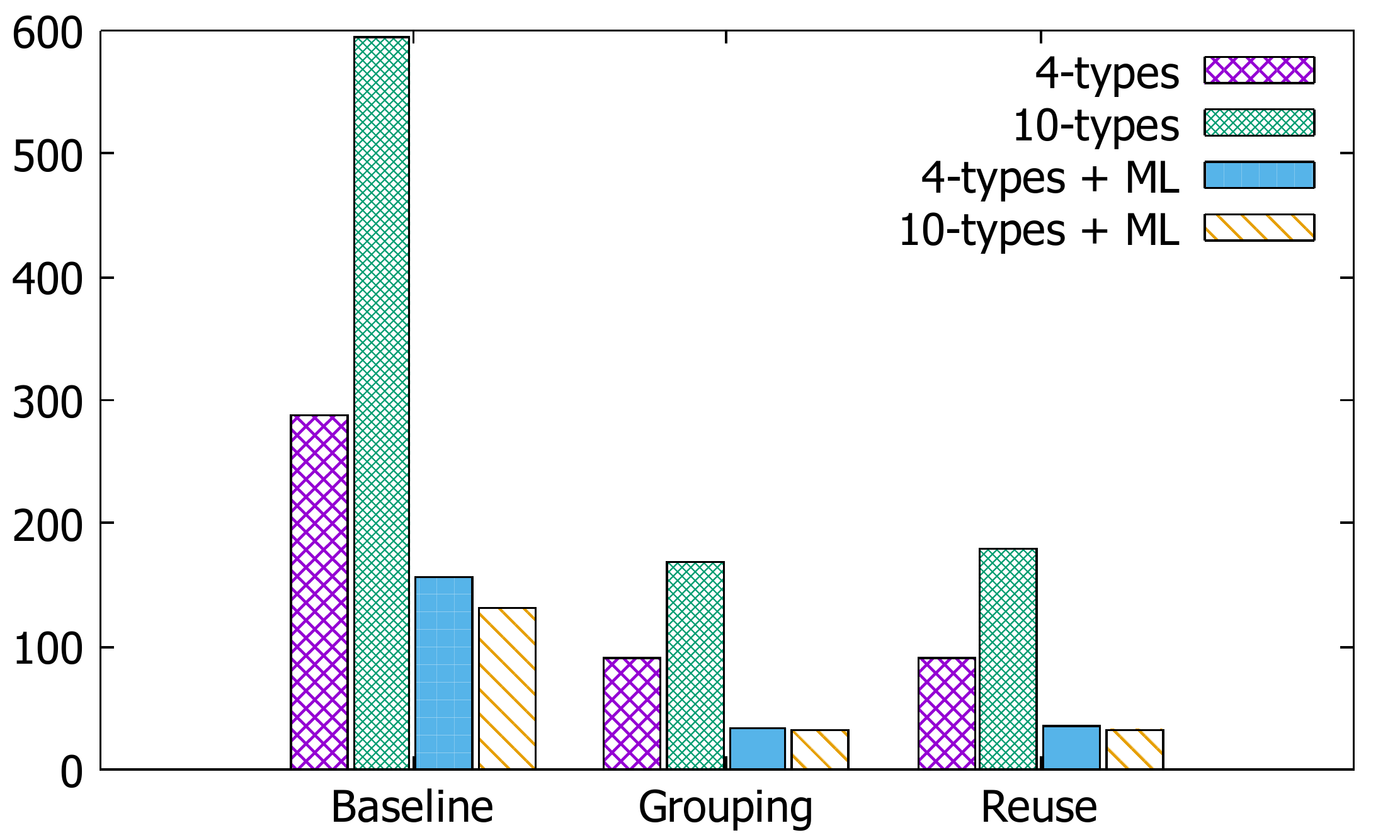}
\caption{\textbf{Execution time for PDF computation
    on a
small workload (6 lines and 3006 points)
with 235 GB input data.} The time unit is second. }
\label{fig:swlET}
\end{figure}

\begin{figure}[t]
\centering
\includegraphics[width=8.5cm]{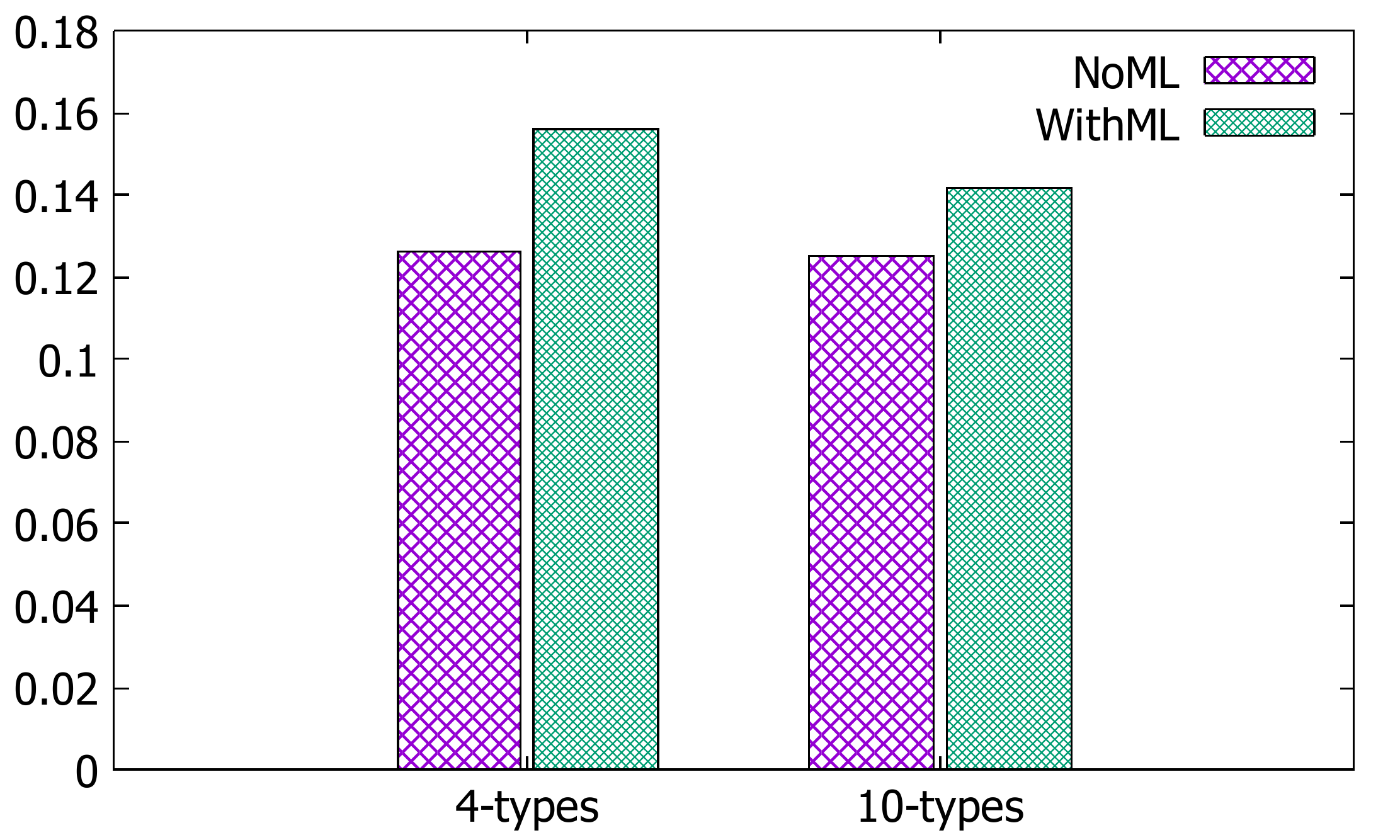}
\caption{\textbf{Error in PDF computation.}
NoML represents the error of different
 methods without adoption of ML, \textit{i.e.}
Baseline, Grouping and Reuse with $4-types$ or $10-types$
(see Figure
\ref{fig:swlET}).
WithML represents the combinations of methods with ML, \textit{i.e.}
Baseline + ML, Grouping + ML, Reuse + ML
with $4-types$ or $10-types$
(see Figure \ref{fig:swlET}).}
\label{fig:swlError}
\end{figure}

In this section, we use a small workload of 6 lines, \textit{i.e.}
3006 points,
to compare the performance of different methods.
The execution time for PDF computation is shown in Figure \ref{fig:swlET} and the error is shown in Figure \ref{fig:swlError}.

We execute the program for the points of the first 6 lines in
Slice 201 using Baseline, Grouping, Reuse, ML and the combination of these
methods with ML.
We take 3 lines as a window for PDF computation
and we process the data of the points in two windows.
The execution time for data loading (see Algorithm
\ref{alg:DL}) is 67s, which is the same for all the methods since
we use the same algorithm. 

Figure \ref{fig:swlET} shows the good performance of our methods:
Grouping (without Reuse), Reuse and ML.
The execution time of Baseline with $10-types$ is much
longer than that with $4-types$. This is expected since the
complexity of  Algorithm \ref{alg:Ana} is $O(n)$, with $n$
distribution types,
and the execution time increases with $n$.
However, with ML, the execution
time is significantly reduced (46\% for
$4-types$ and 78\% for $10-types$).
If we use Grouping without ML or Reuse, the execution time is also reduced a lot
(69\% for $4-types$ and 72\% for $10-types$). This
shows that there are many points that obey the same distribution and
have the same mean and standard deviation values. In addition, since
the data size corresponding to a point is small, \textit{i.e.} 1000 observation values,
the shuffling time for computing the aggregation function is short.
Thus, Grouping outperforms Baseline much. When we couple
Grouping and ML, the advantage becomes more
obvious. The combined method is 88\% and 95\% better compared
with the Baseline for $4-types$ and
$10-types$, \textit{i.e.} the combined method is up to more than
17 times better than Baseline. Reuse is
slightly worse than Grouping, but it is still much
better than Baseline. This is expected since it takes
more time to search for the existing results than to compute PDFs. The
difference between Grouping and Reuse 
is small when the workload is small but it can be significant for
bigger workloads.

Figure \ref{fig:swlError} shows the average error ($E$ in
Equation \ref{eq:averageError}) corresponding to different methods.
In Figure \ref{fig:swlError}, we distinguish between the methods that do not use ML, \textit{i.e.} Baseline, Grouping and Reuse, which we call NoML,
and those that do exploit ML,
\textit{i.e.} Baseline with ML, Grouping with ML and Reuse with ML, which we call WithML.
The methods NoML and WithML have the same error for the same
distribution type set.

The figure shows that $10-types$ does not
reduce much the average error for Baseline (up to
0.0013). However, PDF computation may take more time
when we consider $10-types$ as shown in Figure \ref{fig:swlET}. The average
error is higher for WithML than NoML. The difference is small (up
to 0.017) but WithML can reduce much the execution time of the PDF
computation. In addition, Figure \ref{fig:swlError} shows that
$10-types$ leads to a smaller average error, even though the model
error is higher for WithML. This is reasonable since
there are some types that are very difficult to distinguish in
$10-types$ but the wrong classification of the distribution types
does not increase much the average error. In addition, with more
distribution candidate types, the decision tree produces a better
classification.

\begin{figure}[t]
\centering
\includegraphics[width=8.5cm]{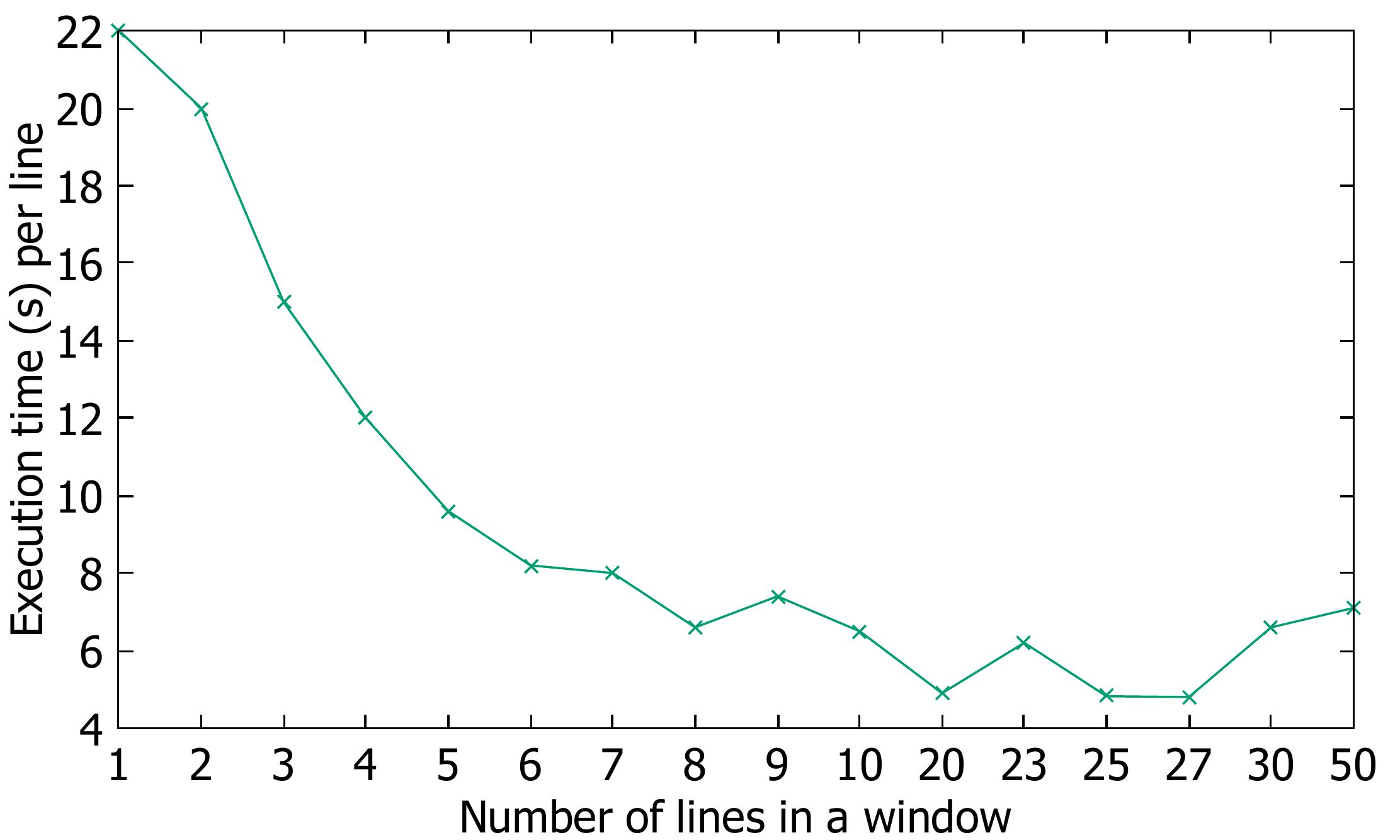}
\caption{\textbf{Average execution time per line for PDF computation with two windows.}}
\label{fig:ws}
\end{figure}

\paragraph{Window Size Adjustment}

The window size is critical for the PDF computation
with Grouping. 
To adjust the window size,
we conduct the following experiment.
We execute the Scala program with Grouping
(with $4-types$ and without ML prediction) for two
windows while each window is composed of different numbers of
lines. Then, we choose a window that corresponds to the shortest
average execution time of each line. Figure \ref{fig:ws} shows the
average execution time of PDF computation for different
window sizes using Grouping. As shown in Figure
\ref{fig:ws}, the average execution time of each line decreases for
larger window sizes. This is reasonable because when the number of
lines increases, more points are aggregated to each group so that
more redundant calculations are avoided. When the window size is 25
lines, the average execution time of each line is minimal. From that point on, when the window size increases, the
execution time also increases. This is expected since when the number
of lines in a window increases, the time to shuffle data among
different nodes increases more than the reduced time by avoiding
redundant calculations. As a result, the average execution time of
each line becomes more significant. However, the average execution
time of data loading stays the same for different window
sizes, \textit{i.e.} about 12s per line.

\begin{figure*}
\centering
\includegraphics{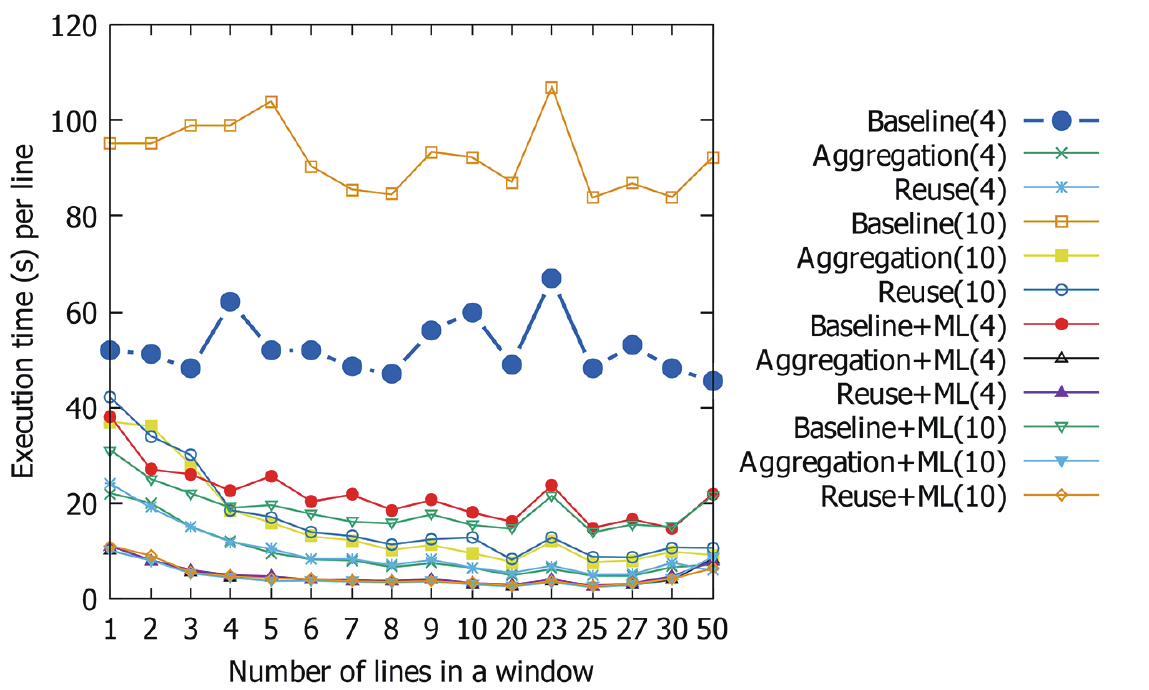}
\caption{\textbf{Average execution time per line for PDF
    computation with two windows.}
With $4-types$ (4) and $10-types$ (10).}
\label{fig:wsm}
\end{figure*}

We measure the execution time of PDF computation
for different window  sizes using other methods. In Figure
\ref{fig:wsm}, we can see that the
window size of 25 is the optimal size for the other methods. In
addition, the execution time of PDF
computation is almost the same for different combinations of
 methods: Grouping plus ML and Reuse plus ML
 both with $4-types$ or $10-types$.
Baseline always
corresponds to longer execution times of PDF computation.
Compared with Baseline, 
Grouping, Reuse and ML can reduce the execution time up to
91\% (more than 10 times faster)
and 84\% (more than 6 times faster). In addition, the
combination of Grouping and ML can be up to 97\%
(more than 33 times) faster. This shows the
obvious performance advantage of our methods in computing
PDFs.

\paragraph{Execution of One Slice}
\label{subsubsec:EOS}

\begin{figure}
\centering
\includegraphics[width=8.5cm]{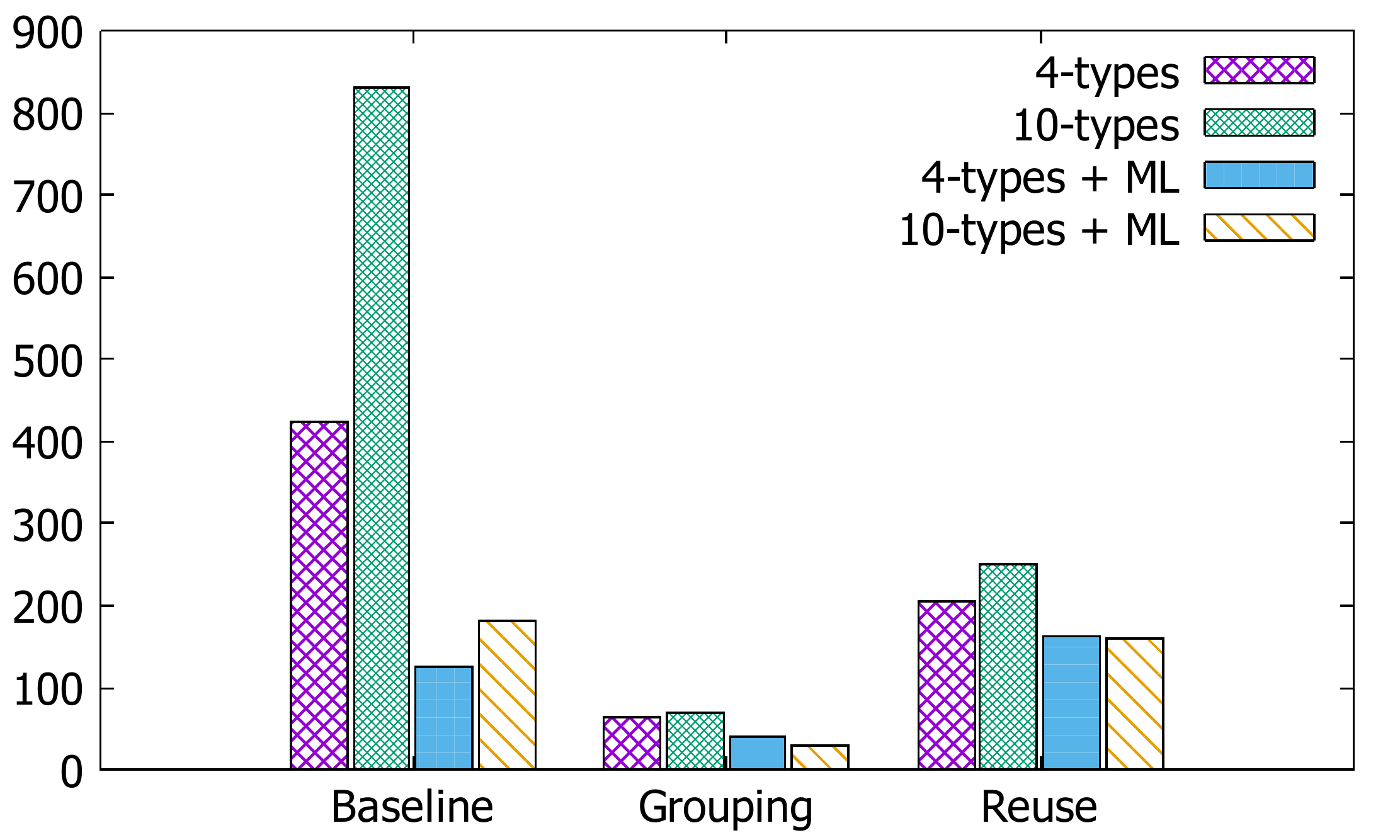}
\caption{\textbf{Execution time of PDF computation of Slice 201 with
    different methods (235 GB input data).} The time unit is minute. The window size is 25 lines.}
\label{fig:25}
\end{figure}

\begin{figure}[t]
\centering
\includegraphics[width=8.5cm]{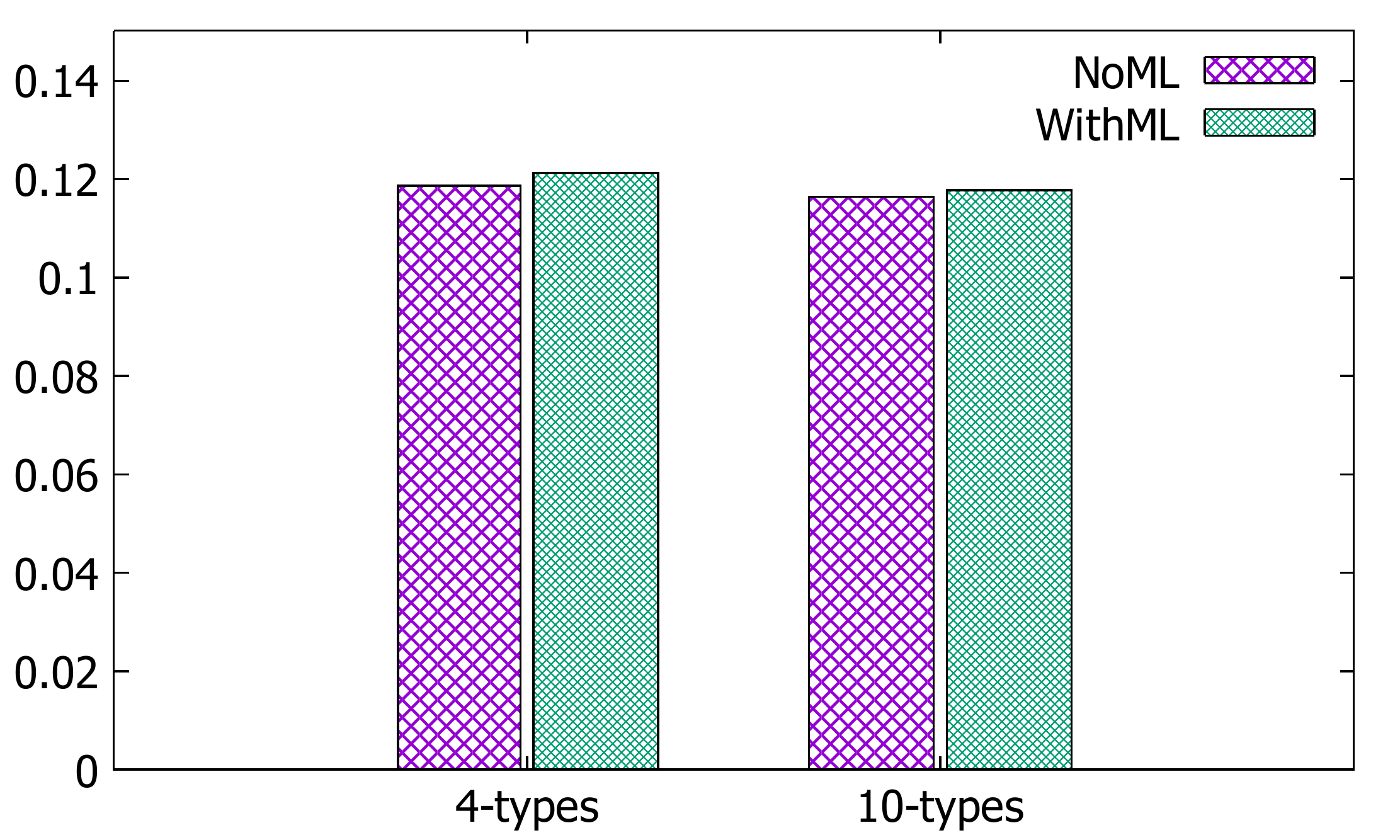}
\caption{\textbf{Error in PDF computation.} NoML stands for Baseline,
  Grouping and Reuse plus
$4-type$ and $10-types$.
WithML stands for Baseline + ML, Grouping + ML, Reuse + ML
$4-type$ and $10-types$.}
\label{fig:25error}
\end{figure}

In this section, we compare the performance of different methods to
execute the program for the points of the whole slice, \textit{i.e.} Slice 201.
We take 25 lines as the window size for PDF computation
and execute the program with different methods the
whole slice, \textit{i.e.} 11 windows of points in Slice 201. The
execution time of data loading is the same for the different
methods, \textit{i.e.} 4098s. The average execution time of each
line is longer than that of the small workload since we cache all the
data in memory during the execution of different methods and
the time to store data increases as the amount of cached data grows. 
The execution time of PDF computation is shown in
Figure \ref{fig:25} and the error is shown in \ref{fig:25error}.

Figure \ref{fig:25} shows that our proposed methods, \textit{i.e.}
Grouping, Reuse and ML, always outperform Baseline for both
$4-types$ and $10-types$. Grouping can reduce
the execution time up to 92\% (more than 10 times) and ML
up to 78\% (more than 3
times faster). The performance of Reuse is better than that of
Baseline when we do not combine ML and the advantage
can be up to 70\% (more than 2 times faster). However, the
performance of the combination of Reuse and ML can be worse
than the combination of Baseline and ML. This is
possible when it takes too much time to search for the existing
results. The combination of Grouping and ML can reduce
the execution time up to 97\% (more than 27 times faster) compared with
Baseline.

Figure \ref{fig:25error} shows the error of PDF computation.
The error is smaller than that of the small workload. The error
of NoML is still smaller than that of WithML. The error for
$4-types$ is almost the same as that for $10-types$. But the error for
$10-types$ using ML is smaller that for $4-types$. This
is similar to what was observed when executing the small
workload. In addition, the error for $10-types$ with ML is
even smaller than the error for $4-types$ without ML. Although there is very small difference (up to 0.0016) of
error between Baseline and ML, the execution
time of the PDF computation process is largely reduce by ML.

\subsubsection{\bf{Scalability Comparisons}}
\label{subsec:CS}

In this section, we study the scalability of Baseline
and the methods that have good performance, \textit{i.e.}
Grouping, ML and Grouping + ML. We use the G5k
cluster with different numbers of nodes from 10 to 60.

\begin{figure}[t]
\centering
\includegraphics[width=8.5cm]{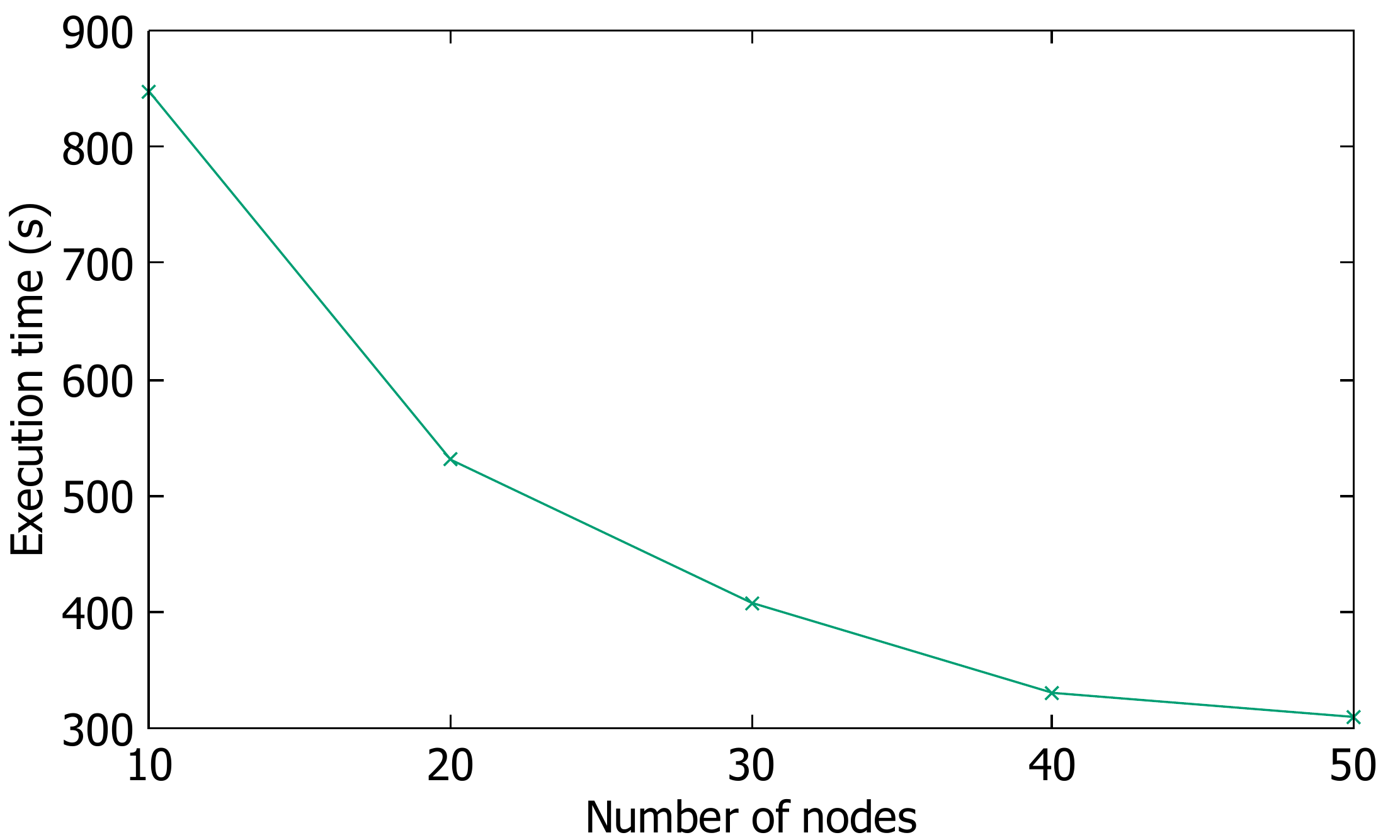}
\caption{\textbf{Execution time for data loading with different numbers of nodes.}}
\label{fig:dl}
\end{figure}

The execution time of data loading with different methods
remains the same for the same number of nodes. Figure \ref{fig:dl} shows
the execution time of data loading, which decreases
rapidly as the number of nodes increases. This indicates the good
scalability of our data loading process.

\begin{figure}[t]
\centering
\includegraphics[width=8.5cm]{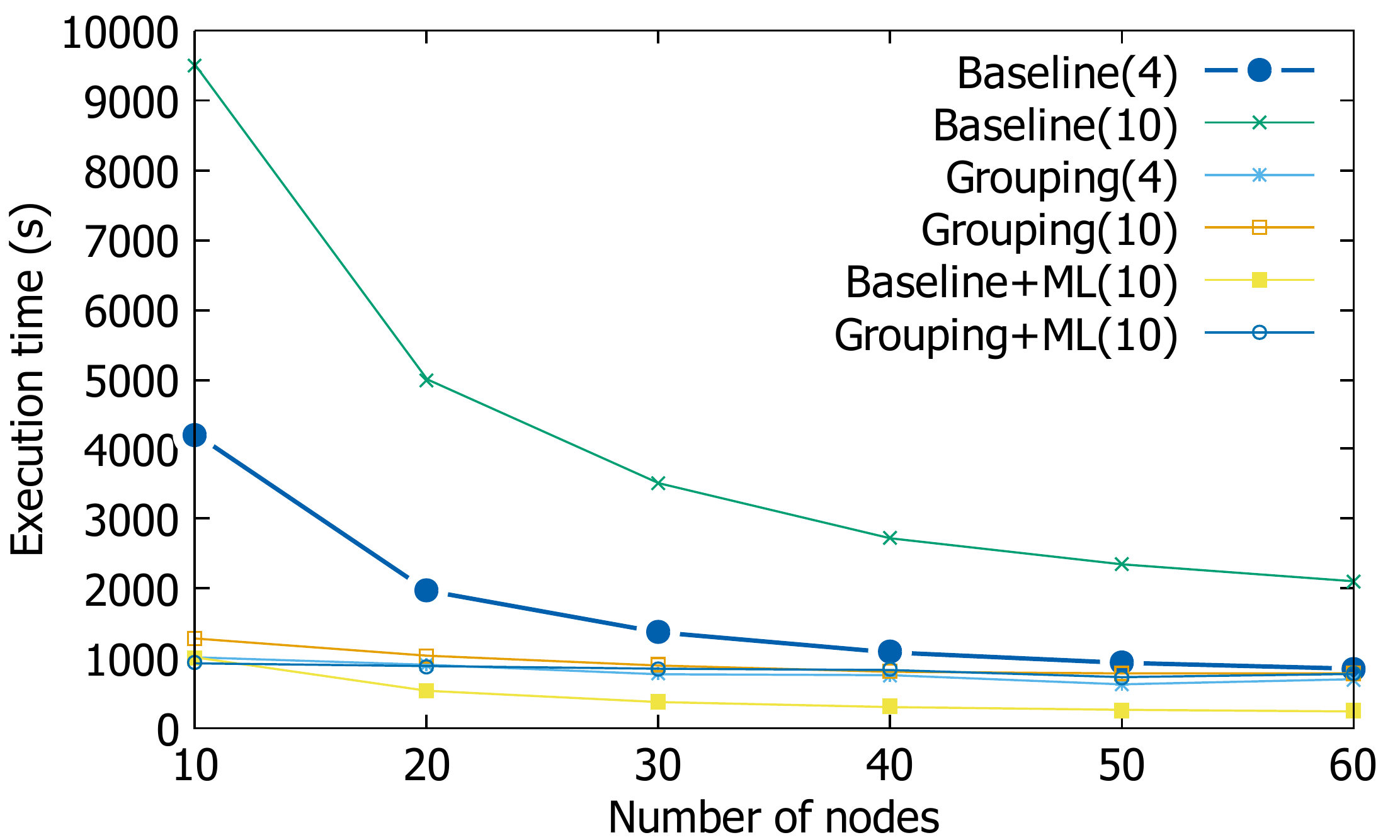}
\caption{\textbf{Execution time of PDF computation with different numbers of nodes.}}
\label{fig:scala}
\end{figure}

The execution time of PDF computation for different
methods and different numbers of nodes is shown in Figure
\ref{fig:scala}. We do not consider Reuse since it is less efficient
than Grouping.
We focus on ML for $10-types$  because it has
small error and an execution time similar to that for $4-types$.
The figure shows that the execution time of
Grouping and ML is always better than that of
Baseline. The advantage of Grouping can be up to
87\% (more than 6 times faster). ML outperforms
Baseline up to 89\% (more than 8 times faster).
Grouping + ML can be better than Baseline by up to 90\%
(more than 9 times faster).
In addition, the execution time of each method decreases as
the number of nodes increases, which indicates that all methods
have good scalability. However, the advantage becomes less obvious
from 50 nodes on.

\begin{figure}[t]
\centering
\includegraphics[width=8.5cm]{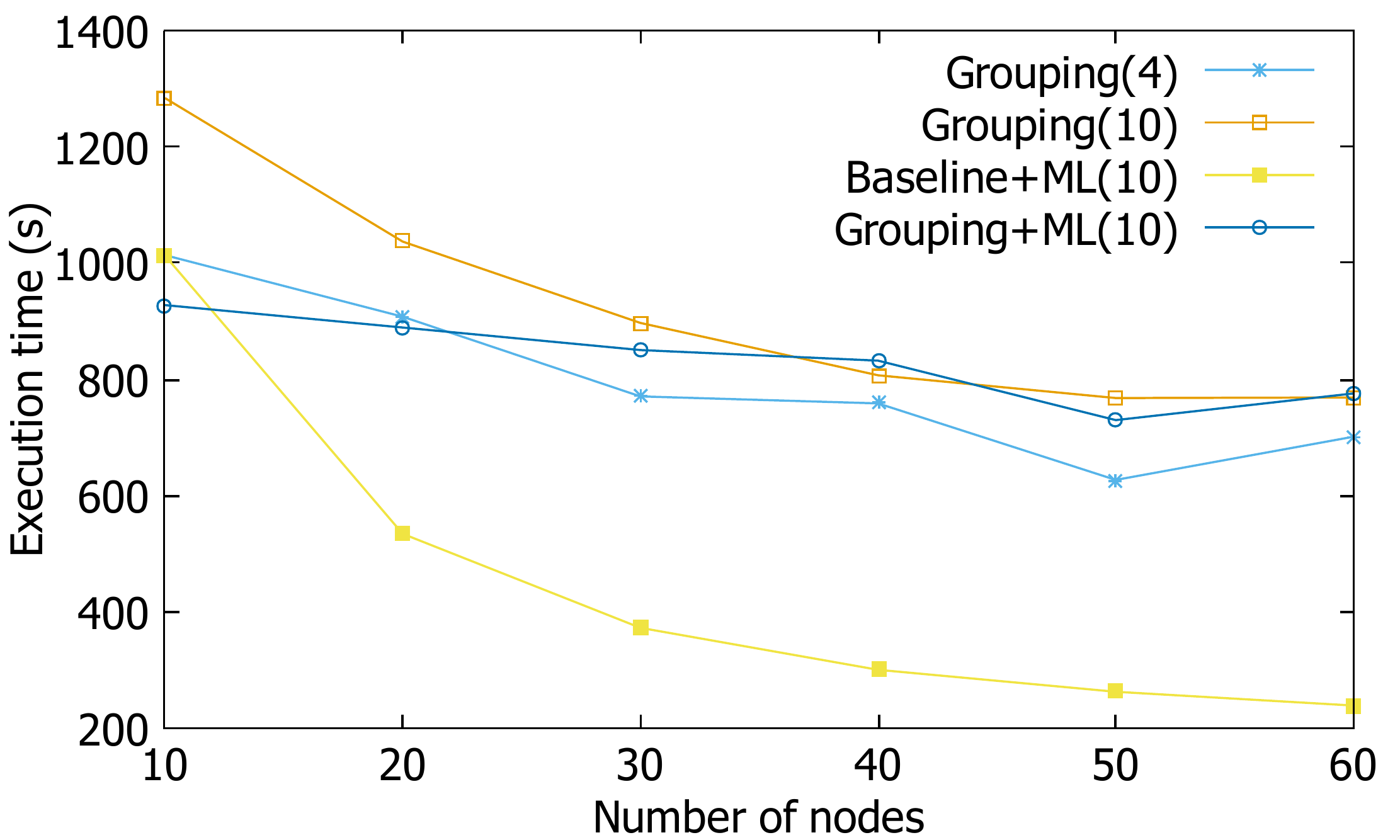}
\caption{\textbf{Execution time of PDF computation with different
    numbers of nodes and a focus on Grouping, ML, Baseline + ML and Grouping + ML.}}
\label{fig:zoom}
\end{figure}

Figure \ref{fig:zoom} gives a focus on our methods.
We do not compare the error of PDF computation since it is similar to that given in Section \ref{subsubsec:EOS}.
The figure shows
that Grouping + ML is better
than either Grouping or ML when the number of nodes is 10.
However, ML starts
outperforming Grouping + ML when the number of nodes exceeds 10.
This is because the shuffling of data among different
nodes takes much time. As the number of nodes increases,
the more time to transfer data among the nodes also increases.
Thus, the
performance of the aggregation function becomes a bottleneck for
Grouping + ML
when the number of nodes is high.

\subsubsection{\bf{Performance of Sampling}}
\label{subsec:EFP}

We now study the efficiency of Sampling. We carried out the experiments in the LNCC cluster. We use two
sampling methods, \textit{i.e.} random sampling and k-means clustering
(see Section \ref{subsec:fproc}). We compare the
performance of the two sampling method with different sampling rates.

\begin{figure}[t]
\centering
\includegraphics[width=8.5cm]{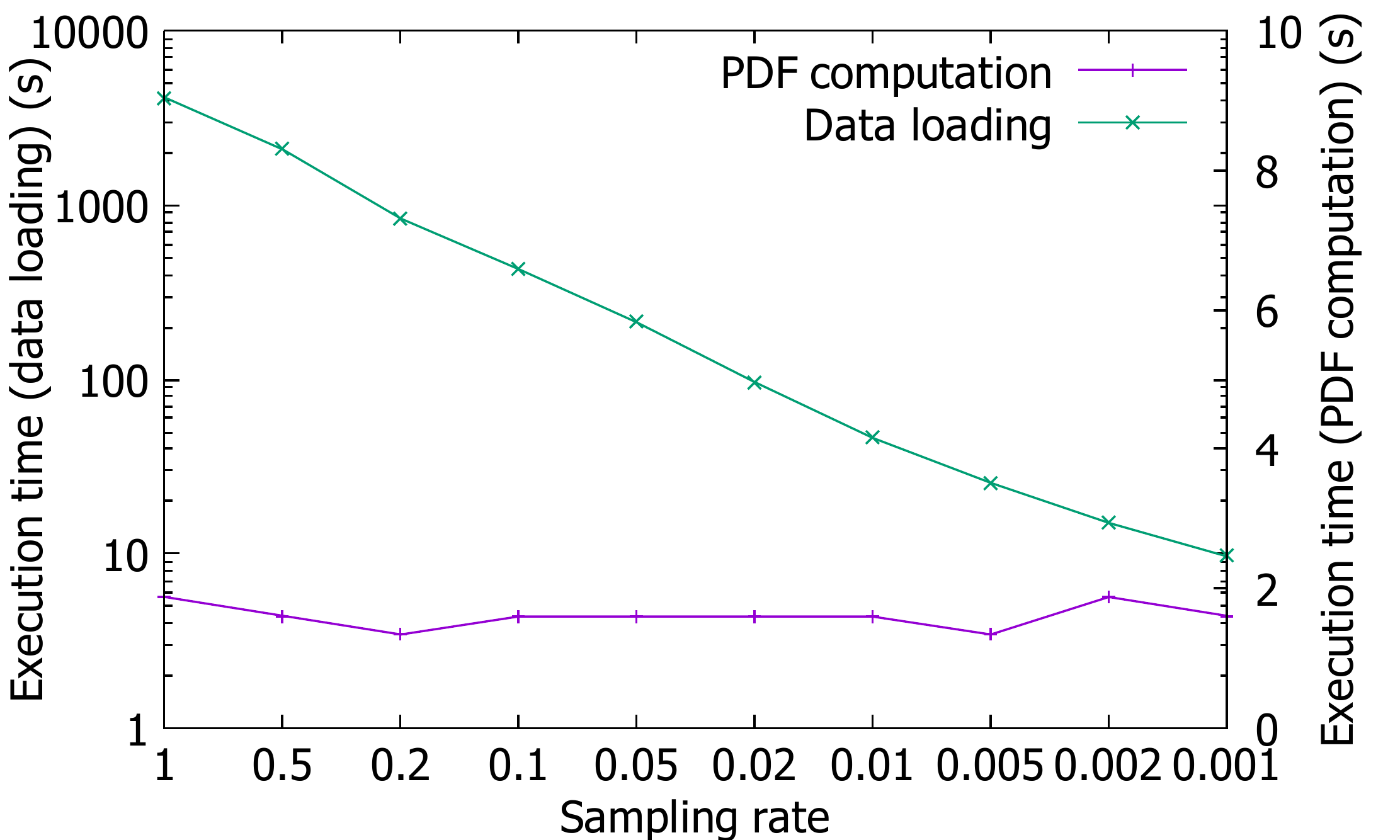}
\caption{\textbf{Execution time with different sampling rates using random sampling.}}
\label{fig:RFP}
\end{figure}

Figure \ref{fig:RFP} shows the execution time of random sampling
with different sampling
rates.
The execution
time for data loading decreases almost linearly as the
sampling rate decreases (both the X and Y axis use a base-10
log scale). This is reasonable since when the sampling rate gets
small, the data loading needs to process less points and thus
loads less data.
The execution time for
PDF computation is very short (about 2 seconds). This is
expected since we avoid calling the R program to compute PDFs and
use a decision tree model to predict the distribution type of each
point. The execution time is almost the same for different
sampling rates since it is already very short and data shuffling
also takes some time. In addition, we only need to transfer the mean
and standard deviation values instead of the whole data set for the
prediction, which also reduces time. This execution time is about 2
seconds, which is very short. However, this method cannot calculate
the error of the PDF computation process. It can help to have a
general view of the whole slice. Then a slice is chosen to compute the
PDFs of the corresponding points.

\begin{figure}[t]
\centering
\includegraphics[width=8.5cm]{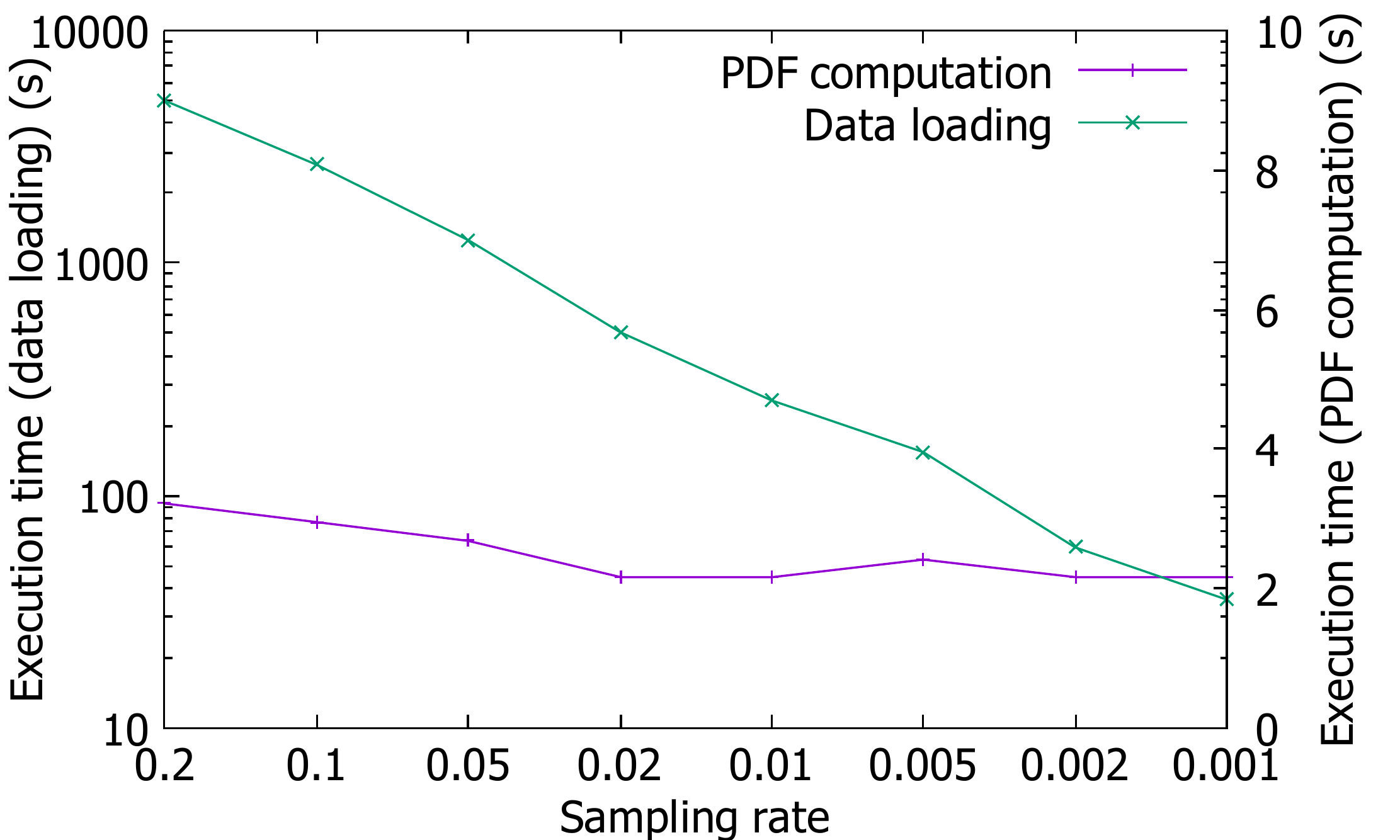}
\caption{\textbf{Execution time with different sampling rates using k-means sampling.}}
\label{fig:kmeans}
\end{figure}

Figure \ref{fig:kmeans} shows the performance of k-means clustering
for sampling the points. The results are almost the same as that of random sampling,
\textit{i.e.} the execution time for data loading decreases
linearly and the execution time for PDF computation is very short
and almost the same for different sampling rates.
The execution time of k-means is longer than that of random for the
same sampling rate. We measure the sampling rate from 0.2 since the
corresponding execution time of k-means for data loading
is already longer than that of random sampling with the sampling
rate of 1.

\begin{figure}[t]
\centering
\includegraphics[width=8.5cm]{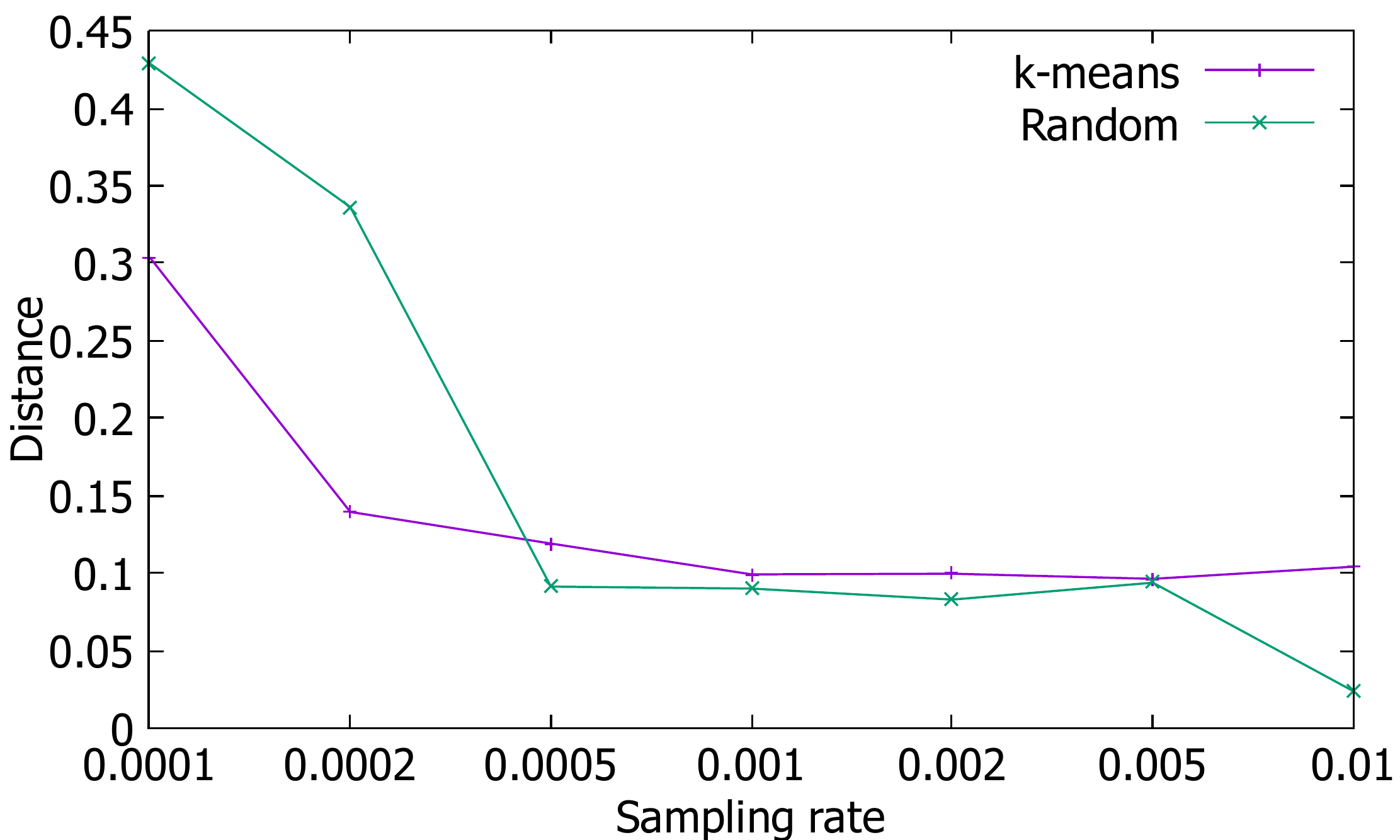}
\caption{\textbf{Distance of the distribution type percentage between the double sampled points and all points.}}
\label{fig:compare}
\end{figure}

Figure \ref{fig:compare} shows the Euclidean distance of the
distribution type percentage between the double sampled points and all
points in one slice using k-means and random sampling.
When the sampling rate is small, the result of k-means is
close to the percentage of overall points. This is because the
double sampled points of k-means method contain diverse
information. However, when the sampling rate is high, the results of
random sampling are similar or better since enough points are
selected with a high sampling rate. Since
random sampling is faster than k-means, we choose it in the
following experiments.

\subsection{\bf{Experiments on Big Data Sets}}
\label{subsec:CTD}

In this section, we compare the performance
of different methods based on big spatial data sets of several TB,
\textit{i.e.} $Set2$ and $Set3$ (see details in Section
\ref{subsec:ES}). We use
the G5k cluster with 30 and 60 nodes.

\subsubsection{\bf{Experiments with 1000 Simulations}} 

In this section, we perform experiments using $Set2$ of
1.9 TB generated by 1000 simulations. We take the
relationship between the combination of mean and standard deviation
values and the distribution types of 25000 points in Slice 0 as
previously generated data to build the decision tree model. The
model error of $4-types$ is 0.02 and the model error of $10-types$ is
0.09. The time to load the model ranges between 1 and 20 seconds,
which is negligible compared with the execution time of PDF
computation.
The time for data loading is 2671 seconds with 30 nodes and 1619
seconds with 60 nodes, which shows the good scalability of the data
loading process.

\begin{figure}[t]
\centering
\includegraphics[width=8.5cm]{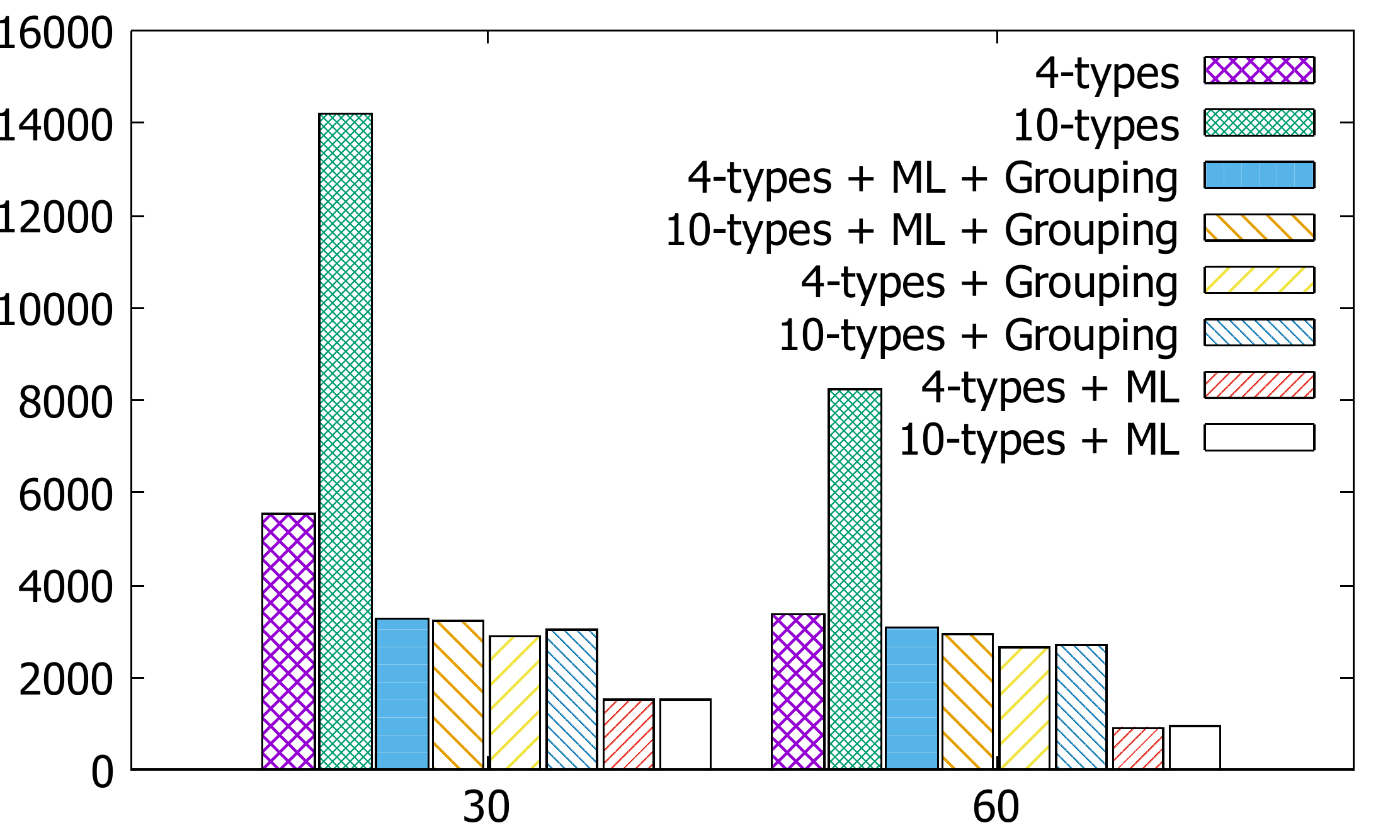}
\caption{\textbf{Execution time for all the points in Slice 201 (1.9 TB input data).} The time unit is second. }
\label{fig:TB18}
\end{figure}

Figure \ref{fig:TB18} shows the good performance of our methods,
\textit{i.e.} Grouping and ML.
Grouping is better than Baseline (up to 79\%) but not as good
as ML, because of data shuffling with many nodes.
And the scalability of Grouping is not good.
ML largely outperforms Baseline (up to 89\%) and has
good scalability.
Because of the shuffling bottleneck, the combination of Grouping and ML
is worse than ML but still better (up to 77\%)
than Baseline. The error of ML with
$10-types$ is always smaller than with
$4-types$.

Finally, we experimented with random sampling to
process the data in two clusters of 30 and 60 nodes.
The execution time of PDF computation ranges between 260s and 280s while
the sampling rate ranges between 0.001 and 1. The average execution
time (272s for 30 nodes and 266 for 60 nodes) is 82\% and 71\% shorter
than the minimum execution time (ML with $4-types$ for 30 and 60
nodes) shown in Figure \ref{fig:TB18}.
Note that doubling the number of nodes does not yield much
improvement. This is because using more
nodes increases data transfers to send the mean and standard
deviation values and the distribution type from each node to the
Spark master node in Spark cluster to compute the distribution
percentage.

\subsubsection{\bf{Experiments with 10000 Simulations}} 

\begin{figure}[t]
\centering
\includegraphics[width=8.5cm]{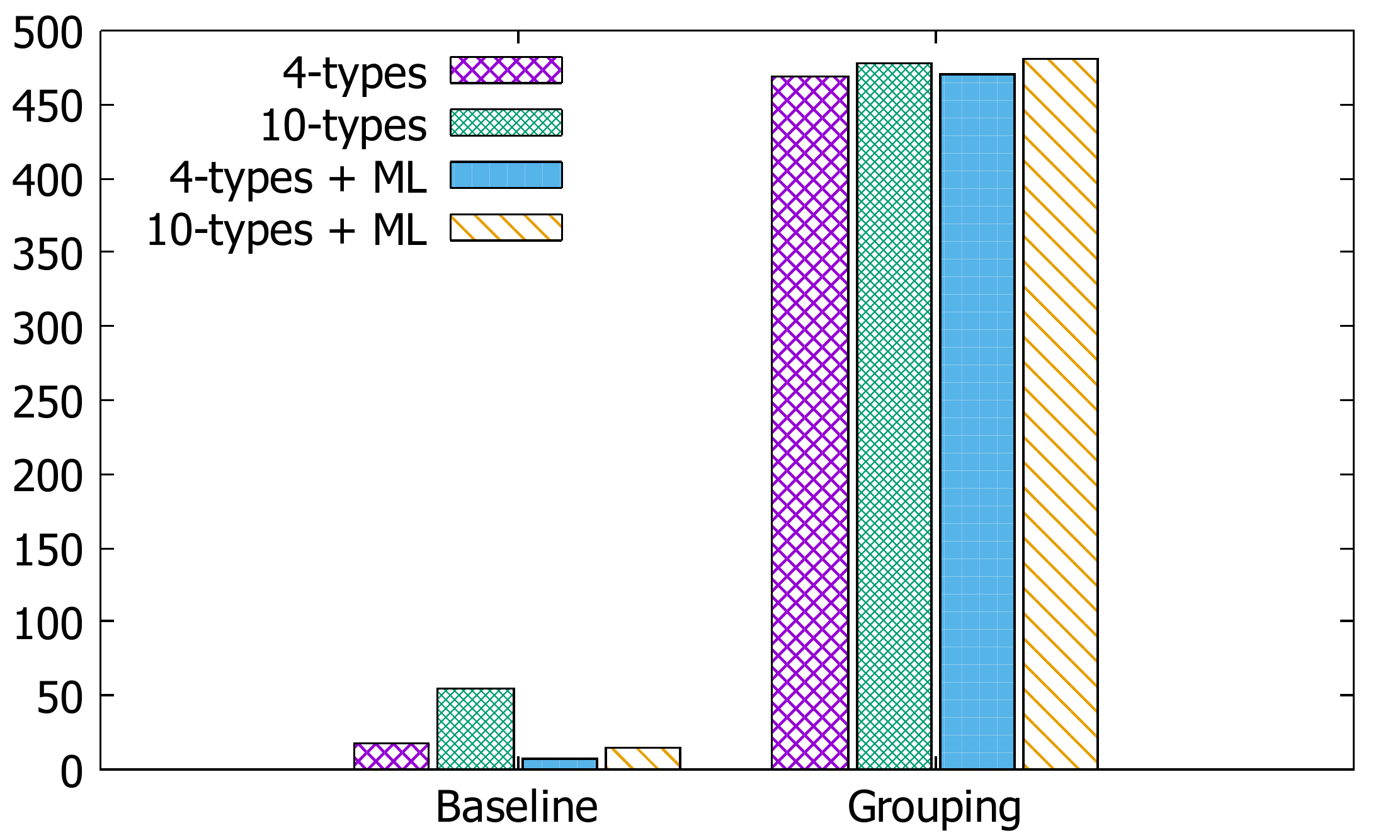}
\caption{\textbf{Execution time for PDF computation on small workload
    (2 lines and 1002 points) with 2.4 TB input data.} The time unit is second. }
\label{fig:TBS}
\end{figure}

In this section, we perform experiments using $Set3$ of 2.4 TB
generated by 10000 simulations.
Again, we take the relationship between the combination of
mean and standard deviation values and the distribution types of 25000
points in Slice 0 as the previously generated data for the
decision tree model. The model error of $4-types$ is 0.006 and the
model error of $10-types$ is 0.012. The time to load the model ranges
between 1 and 20 seconds, which is negligible compared with the
execution time of PDF computation.
First, we execute the program for the points of the first 2 lines in
Slice 201 in a cluster of 30 nodes. We take 1 line as a window, and we
process the data of two windows. The execution time for data
loading (Algorithm \ref{alg:DL}) is 28s, which is the same for
all the methods since we use the same algorithm (Algorithm
\ref{alg:DL}). The execution time for PDF computation is
shown in Figure \ref{fig:TBS}.

Figure \ref{fig:TBS} shows the superior performance of ML. The
execution time of Baseline with $10-types$ is much longer than that of
$4-types$ as expected.
Compared with Baseline, ML reduces execution time much (57\% with $4-types$ and 72\% with $10-types$).
However, the execution time of Grouping is much longer. This is
because the data size of each point is 9 times bigger since a
point corresponds to 10000 observation values instead of 1000. During
Grouping, much more data is transferred among different
nodes, which takes much time. Thus, in the next experiment, we do not
use Grouping.

We also measure the error ($E$ defined in Equation
\ref{eq:averageError}) during execution.
$10-types$ does
not reduce much the average error for Baseline (up to
0.008). However, it may take much more time for PDF
computation  when we consider $10-types$ as shown in Figure
\ref{fig:TBS}. Using WithML, the average error is
slightly bigger than that of NoML. The difference is very small (up to
0.007) but ML can reduce much the execution time of PDF
computation. Furthermore, $10-types$ leads to smaller average
error even though the model error is bigger when using ML.

\begin{figure}[t]
\centering
\includegraphics[width=8.5cm]{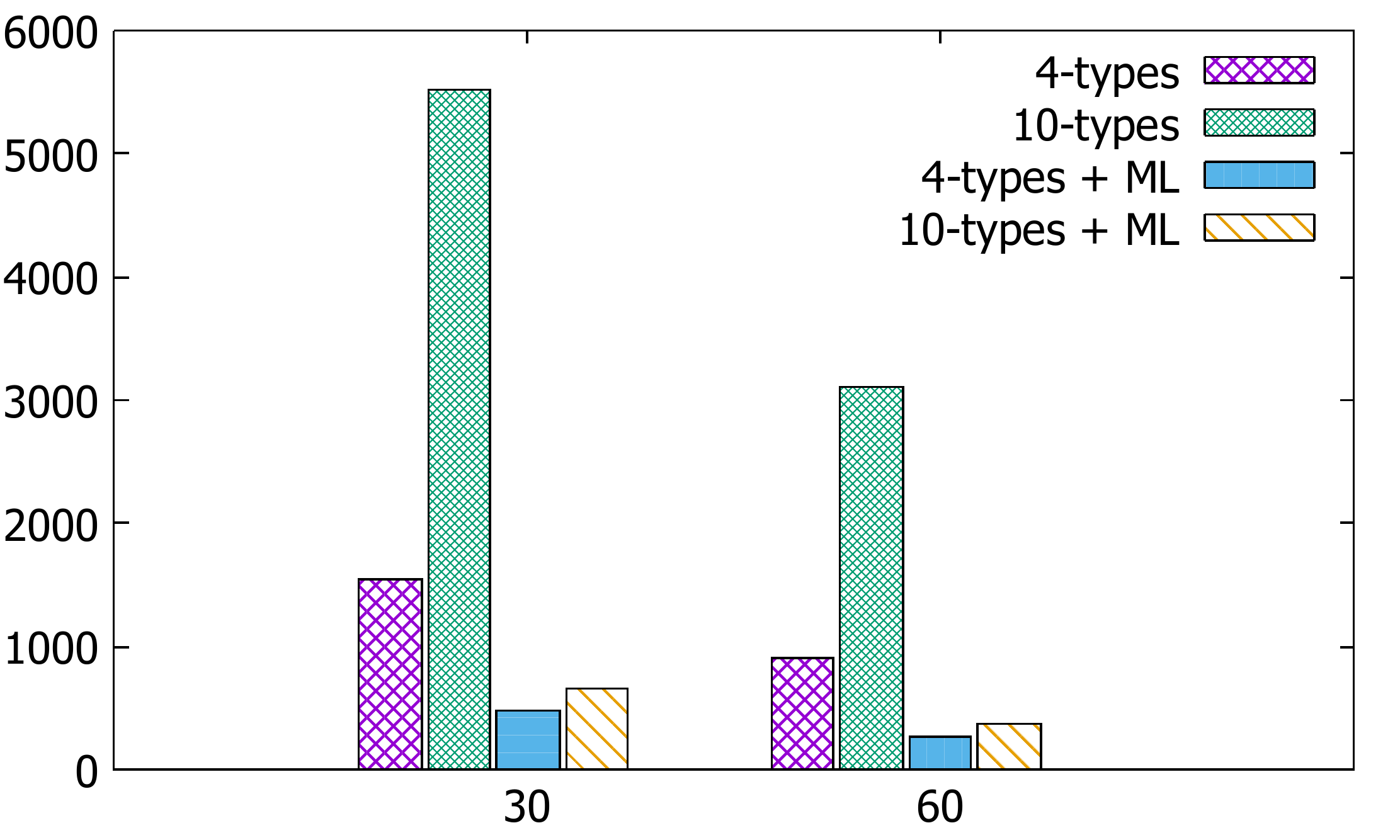}
\caption{\textbf{Execution time for all the points in Slice 201 (2.4 TB input data).} The time unit is second. }
\label{fig:TBA}
\end{figure}

Now, we execute the program for all points in Slice 201. As we only
compare the execution with Baseline and ML, we take
126 lines as a window in order to parallelize the execution of
different points in different nodes.
The time for data loading is 4592 seconds with 30 nodes.
Figure \ref{fig:TBA} shows the superior performance of ML (up to 88\%,
\textit{i.e.} more than 7 times faster
than Baseline
).
Furthermore, the PDF computation process scales very well.
We measured the average error and found that ML
incurs small error while reducing
execution time much. Finally, we use the random sampling
to process the data in two clusters of 30 and 60 nodes.
The
execution time of PDF computation ranges between 128s and
200s while the sampling rate ranges between 0.001 and 1. The average
execution time (172s for 30 nodes and
155s for 60 nodes) is 64\% and
41\% smaller than the minimum execution time (ML with $4-types$ for 30
and 60 nodes) shown in Figure \ref{fig:TBA}.
Again (as for the experiment with 1000 simulations), doubling the
number of nodes does not yield much improvement.

\section{Conclusion}\label{sec:con}

Uncertainty quantification of spatial data 
requires computing a Probability Density Function (PDF) of each point
in a spatial cube area. 
However, computing PDFs on big spatial data, as produced by applications in scientific areas such as geological or seismic interpretation,  
can be very time consuming.

In this paper, we addressed the problem of efficiently computing PDFs under bounded error constraints.
We proposed a parallel solution using a Spark cluster
with three new methods: Grouping, ML and Sampling.
Grouping aggregates the points of the same
statistical features together in order to reduce redundant
calculation. This method is very efficient when the data to be
transferred is not too big and the number of cluster nodes is
small. ML generates a decision tree model based
on previously generated data and predicts the distribution type of a
point in order to avoid useless calculation based on wrong
distribution types.
Sampling  enables to efficiently compute statistical parameters of a region by sampling a fraction of the total number of points to reduce the computation space.

To validate our solution, we implemented these methods in a Spark cluster and performed extensive experiments on 
two different computer clusters (with 6 and 64 nodes)
using big spatial data ranging from hundreds of GB to several TB.
This data was generated from simulations
based on the models from a seismic benchmark for oil and gas exploration,
which includes models for seismic wave propagation.

The experimental results show that our solution is efficient and scales up very well compared with Baseline.
Grouping outperforms Baseline by up to 92\% (more than 10 times) without introducing extra error.
ML can be up to 91\% (more than 9 times)
better than Baseline with very slight acceptable error (up to 0.017). The combination of Grouping and ML can be up to 97\% (more than 33 times) better than Baseline. 
As the number of nodes exceeds 10 nodes,
ML outperforms the combination.
Thus, in order to compute PDFs, the combination of Grouping and ML is
the optimal method when each point corresponds to a small number of observation values, e.g. 1000, and when there is small number of nodes (less than 20).
Otherwise, ML is the best option.
We also showed that Sampling is very efficient to calculate general
statistics information in order to choose a slice for calculating
PDFs.
Finally, Sampling should be used with the aforementioned best option in order to efficiently compute PDFs.

\section*{Acknowledgment}
This work was partially funded by EU H2020 Project HPC4e
with MCTI/RNP-Brazil, CNPq, FAPERJ, and Inria Associated Team SciDISC.
The work of J. Liu, E. Pacitti and P. Valduriez were performed in the
context of the Computational Biology Institute (\url{http://www.ibc-montpellier.fr}).
The  experiments were carried out using a cluster at LNCC in Brazil and
the Grid5000 testbed in France (\url{https://www.grid5000.fr}).

%
%

\bibliographystyle{abbrv}
\balance
\bibliography{biblio}

\end{document}